\documentstyle[aps,prb,epsf]{revtex} 
\begin{document}
\title{ The effects of weak disorders on Quantum Hall critical points }
\author{Jinwu Ye }
\address{Department of Physics and Astronomy,
The Johns Hopkins University, Baltimore MD, 21218}
\date{\today}
\maketitle
\begin{abstract}
We study the consequences of random mass, random scalar potential
and random vector potential on the line of fixed points
between integer/fractional quantum Hall states
and an insulator. This line of fixed points was first identified
in a clean Dirac fermion system with both Chern-Simon coupling and Coulomb
interaction in Phys. Rev. Lett. {\bf 80}, 5409 (1998).  
  By performing a Renormalization Group analysis
  in $ 1/N $ ( $ N $ is the No.  of species of Dirac fermions) 
  and the variances of three disorders
  $ \Delta_{M}, \Delta_{V}, \Delta_{A} $,
  we find that $ \Delta_{M} $ is irrelevant along this line,
  both $ \Delta_{A} $ and $ \Delta_{V} $ are marginal.
  With the presence of all the three disorders, the pure fixed line
  is unstable.
  Setting Chern-Simon interaction to zero, 
  we find one non-trivial lines of fixed points in $ (\Delta_{A}, w) $ plane
  with dynamic exponent $ z=1 $ and continuously changing $ \nu $,
  it is {\em stable} against small $ (\Delta_{M},\Delta_{V} ) $ 
  in a small range of the line $ 1< w < 1.31 $, therefore it may
  be relevant to integer quantum Hall transition.
  Setting $ \Delta_{M} =0 $, we find a fixed plane with $ z=1 $,
  the part of this plane with $ \nu > 1 $ is stable against small
  $ \Delta_{M} $, therefore it may be relevant to fractional
  quantum Hall transition.
  Although we do not find a generic 
  fixed point with all the couplings {\em non-vanishing}, we prove that
  the theory is {\em renormalizable } to the order $ (1/N)^{2},
  (1/N) \Delta, \Delta^{2} $ and explore the
  interesting processes which describe the interferences between
  Chern-Simon interaction, Coulomb interaction and the three kinds
  of disorders.

\end{abstract}
%\narrowtext

\section{Introduction}
The zero temperature quantum phase transitions between the different quantum Hall
and insulating states of a two-dimensional electron gas in a strong magnetic field 
are among the most intensively studied quantum critical points, 
both theoretically~\cite{bodo,shahar} and experimentally~\cite{expts}.
Earlier theoretical investigations focussed on the transitions
between the Integer Quantum Hall plateaus and 
described them in terms of non-interacting electrons moving 
in a random external potential~\cite{ammp}.

  Ludwig {\sl et. al} introduced and analyzed a Dirac fermion model
  with random mass, random scalar potential and random gauge potential.
  They found that the random mass is marginally
   {\em irrelevant}, random scalar potential marginally {\em relevant}.
  However, they found that random vector potential is exactly {\em marginal},
  therefore there is a line of fixed points characterized by the strength
  of the random gauge potential, the zero energy wavefunction shows
  multifractal behaviors with the exponents continuously changing along
  the line\cite{lud}.  If two of the
   random potentials are non-zero, the third one will be generated and
   the system flows to strong coupling regime. They further argued 
   that the system should flow to the generic fixed point
   of integer Quantum Hall transitions which was correctly described
   by Chalker-Coddington model \cite{cha,kiv}. 
  
  The properties of the fixed line of Dirac fermions in the presence of 
  random magnetic fields were further investigated
  in Ref.\cite{chamon,liou}. Later, the $ U(1) $ gauge potential was extended
  to non-Abelian gauge potential \cite{non1,non2,herm}.

It has also been argued that the transitions between fractional quantum Hall
states could be mapped onto models essentially equivalent
to those between the integer states~\cite{jkt}. The latter point of
view was however questioned by Wen and Wu~\cite{wen} and
Chen, Fisher and Wu~\cite{chen}:
they focussed on the simpler case of systems in the presence of
a {\em periodic} rather than a random potential, and examined a model
of anyons, with a statistical angle $\theta$ and short-range repulsive
interactions, which displayed a second order quantum phase transition
between a quantized Hall state and a Mott insulator as the strength of the
periodic potential was varied.
This transition was characterized by a line of critical points
with continuously varying exponents, parameterized by the value
of $\theta$.
For the case $\theta = 0$, when the anyons were fermions, 
the transition was out of a integer quantum Hall state; its exponents and
other universal properties were different from the cases
$0 < \theta < 2 \pi$ for which the anyons acquired fractional statistics
and the transition was out a fractional quantum Hall state. 
(For $\theta = 2 \pi$ the anyons became bosons and the Hall
state reduced to a superfluid.)

In all of the above theoretical works, the long-range 
Coulomb interactions between charge carriers have been effectively ignored
However, a few recent works have taken steps to remedy this
serious shortcoming. Yang {\em et al.}~\cite{yang} studied the integer
quantum Hall transition under a Hatree-Fock treatment of the Coulomb
interaction. Lee and Wang~\cite{lee} showed that the renormalization
group eigenvalue of the Coulomb interaction was zero at the Hartree-Fock 
critical point; higher order calculations are therefore
necessary to understand the physics.
Pfannkuche and MacDonald~\cite{allan} numerically studied
electrons with Coulomb interactions in a periodic potential 
between a fractional Hall state
and an insulator, but were limited to rather small system sizes.
Interesting scaling interpretations of Coulomb interaction-induced
dephasing were discussed in Ref~\cite{polya}.

  Most recently, neglecting the disorders,
  Ye and Sachdev provided a thorough analysis of the consequences
of Coulomb interactions on the anyons in a periodic potential model
of Refs~\cite{wen,chen}. They showed that the Coulomb interaction
is {\em marginally irrelevant} for the integer case ($\theta=0$), and remains
so for the fractional case for small values of $\theta$; this marginally irrelevant
interaction will lead to logarithmic corrections to naive scaling functions for
the vicinity of the transition.
For larger $\theta$, they established, in a certain $1/N$ expansion, the
existence of a novel line of fixed points at which the Coulomb interactions
acquire a non-zero fixed point value determined by the value of $\theta$. 
There are no logarithmic corrections at these fixed points, and naive scaling holds. They found a dynamic critical exponent $z=1$ at all points on the fixed line, providing
a concrete realization of the scenario~\cite{mpaf,shahar}, 
not previously established explicitly,
that energies must scale as inverse distances for the $1/r$ Coulomb interaction.
 They also found the correlation length exponent $ \nu > 2/d $ ( where $ d $
 is the spatial dimensionality ) along this fixed line, which implies
 that the fixed line is stable against weak random mass disorder.

    In this paper, we study in detail the consequences of random mass, 
    random scalar potential and random vector potential at
    the line of fixed points. In the rest of this section, we introduce
    the notations and the model. 
    In the next section, we give very general Renormalization Group (RG0
    formulation of the model and establish some {\em exact} results
    ( for example, Eqs.\ref{ward} and \ref{zbeta} ). In Sec.III, we
    perform one-loop expansion and discuss the implications of R.G.
    flow equations. In Sec.IV, we first discuss the R.G. equation
    at $ N=\infty $ limit, then calculate $ 1/N $ corrections and
    discuss the solutions in different cases. Finally, we reach conclusions
    in Sec. V. In the appendix, we show the equivalence of the two
    forms of the random vector gauge potential.

We begin our analysis by writing down the model of Ref~\cite{jinwu}
extended to include random mass $  M(x) $,
random scalar potential $ V(x) $ and random vector potential 
$ A_{i}(x) $ \cite{lud}.

\begin{eqnarray}
 {\cal S} & = & \int d^d x d \tau \Biggl[ \alpha  \overline{\psi}_{m} \gamma_{0} \partial_{0}
    \psi_{m} + \overline{\psi}_{m} \gamma_{i} \partial_{i} \psi_{m} 
  -\frac{i}{\sqrt{N}} q \mu^{\epsilon/2} \alpha^{1/2} a_{0}
\overline{\psi}_{m} \gamma_{0} \psi_{m}
-\frac{i}{\sqrt{N}} g \mu^{\epsilon/2} \alpha^{1/2} a_{i}
\overline{\psi}_{m} \gamma_{i} \psi_{m} \Biggr]      \nonumber\\
    & + & \int \frac{d^2 k}{4 \pi^2} \frac{d \omega}{2 \pi}
\left[i k a_{0}(-\vec{k},-\omega) a_{t}(\vec{k}, \omega)
+ \frac{k}{2} a_{t}(-\vec{k},-\omega) a_{t}(\vec{k}, \omega) \right]
		\nonumber  \\
  & + & \int d^d x d \tau \Biggl[ i M(x) \overline{\psi}_{m} \psi_{m} 
  + V(x) \overline{\psi}_{m} \gamma_{0}  \psi_{m}
  + i A_{i}(x) \overline{\psi}_{m} \gamma_{i} \psi_{m} \Biggr] 
\label{classical}
\end{eqnarray}
The $\psi_m$ are $m=1\cdots N$ species of charge $q/\sqrt{N}$
2+1 dimensional Dirac fermions which interact with a $U(1)$ gauge field
$(a_0, a_i)$ ($i=1,2$); we are interested in the case $N=1$ but will find the
large $N$ expansion to be a useful tool. The $\gamma_0, \gamma_i$ are the Dirac
$\gamma$ matrices, $x_i$ ($\tau$) are spatial (temporal) co-ordinates
with $\partial_0 \equiv \partial_{\tau}$, $\partial_i \equiv \partial_{x_i}$,
and $\vec{k}$, $\omega$ ($k = |\vec{k}|$)
are the Fourier transformed wavevector and frequency
variables.
To aid the subsequent renormalization group analysis, we are working in $d=2+\epsilon$
spatial dimensions and $\mu$ is a renormalization scale. 
The parameter $\alpha$ is introduced to allow for anisotropic renormalization
between space and time~\cite{cardy}.
We have used the Coulomb
gauge which allows us to explicitly represent $a_i$ in terms of the
transverse spatial component with $a_i = i \epsilon_{ij} k_j a_t / k$.
The $ a_{0} a_{t} $ term in ${\cal S}$ is the Chern Simons coupling:
it turns the Dirac  particles into anyons with a statistical angle
$\theta/N$ with $\theta \equiv qg$; notice
 that the angle is of order $1/N$ and so the expected periodicity of the
physics under $\theta/N \rightarrow \theta/N +4\pi$ will not be visible in the $1/N$ expansion. 
The $ a_{t} a_{t} $ term is the Coulomb interaction,
and it has been written in terms of $a_t$ following Ref~\cite{bert}.

In the absence of the Coulomb interaction and the random potential terms,
it was shown in 
Ref~\cite{chen} that ${\cal S}$ represents the critical theory of a system
of anyons in a periodic potential undergoing a transition from an insulator
with conductivities $\sigma_{xx} = \sigma_{xy} = 0$ into a fractional
quantum Hall state with $\sigma_{xx} = 0$ and $\sigma_{xy} = (q^2 /h)/(1- \theta/2 \pi)$
to leading order in $1/N$.
Both these states have energy gaps. It was shown 
in Ref.\cite{jinwu} that the Coulomb
interaction and disorder do not modify the values of
$\sigma_{ij}$ in either phase.

The relationship of the continuum model ${\cal S}$ to the more realistic model
of electrons studied in Ref~\cite{allan} remains somewhat unclear,
although it is 
plausible that ${\cal S}$ is the critical theory of the latter. We may also 
view ${\cal S}$ as the simplest theory consistent with the following requirements,
and therefore worthy of further study: ({\em i\/}) the two phases on either side of
the critical point have the correct values of $\sigma_{ij}$, and the Hall phase
has {\em both\/} quasi-particle and quasi-hole excitations with the correct
charge and statistics, and
({\em ii\/}) the gap towards the quasi-particle {\em and\/} the quasi-hole
excitations vanishes at the critical point.

\section{The Renormalization Group Formulation of the Model }

We now proceed with a renormalization group analysis of ${\cal S}$. 
Simple power counting shows that the Chern-Simons, 
Coulomb interactions~\cite{subir} and all three kinds of disorders
are marginal at tree level in $d=2$, and so loop expansions are required 
and useful. Power counting also shows that a short-range four-fermion interaction
term is {\em irrelevant} and has therefore been neglected in ${\cal S}$; this makes
the fermionic formulation of the anyon problem much simpler than its
bosonic counterpart~\cite{wen,fisher,boson}.

   We assume all the three kinds of disorder satisfy Gaussian distribution
   with zero mean and variances $ \Delta_{M}, \Delta_{V}, \Delta_{A} $.
\begin{eqnarray}
    <M(x) M(x^{\prime} ) > & = & \Delta_{M} \delta^{d}(x-x^{\prime}) 
	       \nonumber  \\
    <V(x) V(x^{\prime} ) > & = & \Delta_{V} \delta^{d}(x-x^{\prime}) 
		\nonumber  \\
    <A_{i}(x) A_{j}(x^{\prime} ) >  & = & \delta_{ij} \Delta_{A}
    \delta^{d}(x-x^{\prime})
\label{aver}
\end{eqnarray}

    By introducing replica $ a, b=1,2, \cdots, n $ and doing quenched average
    over the above Gaussian distributions, we get
\begin{eqnarray}
 {\cal S} & = & \int d^d x d \tau \Biggl[ \alpha 
  \overline{\psi}^{a}_{m} \gamma_{0} \partial_{0} \psi^{a}_{m}
  + \overline{\psi}^{a}_{m} \gamma_{i} \partial_{i} \psi^{a}_{m} 
  -\frac{i}{\sqrt{N}} q \mu^{\epsilon/2} \alpha^{1/2} a^{a}_{0}
\overline{\psi}^{a}_{m} \gamma_{0} \psi^{a}_{m}
-\frac{i}{\sqrt{N}} g \mu^{\epsilon/2} \alpha^{1/2} a^{a}_{i}
\overline{\psi}^{a}_{m} \gamma_{i} \psi^{a}_{m} \Biggr]      \nonumber\\
    & + & \int \frac{d^2 k}{4 \pi^2} \frac{d \omega}{2 \pi}
\left[i k a^{a}_{0}(-\vec{k},-\omega) a^{a}_{t}(\vec{k}, \omega)
+ \frac{k}{2} a^{a}_{t}(-\vec{k},-\omega) a^{a}_{t}(\vec{k}, \omega) \right]
		\nonumber  \\
  & + & \int d^d x d \tau d \tau^{\prime}
  \Biggl[ \Delta_{M} \mu^{\epsilon}
  [\overline{\psi}^{a}_{m}(x,\tau) \psi^{a}_{m}(x,\tau)] 
   [\overline{\psi}^{b}_{m}(x,\tau^{\prime}) \psi^{b}_{m}(x,\tau^{\prime})] 
 -\Delta_{V} \mu^{\epsilon}
  [\overline{\psi}^{a}_{m}(x,\tau) \gamma_{0} \psi^{a}_{m}(x,\tau)] 
  [\overline{\psi}^{b}_{m}(x,\tau^{\prime}) \gamma_{0} 
  \psi^{b}_{m}(x,\tau^{\prime})]   \nonumber   \\
  & & ~~~~~~~~~~~~~~+\Delta_{A} \mu^{\epsilon} [\overline{\psi}^{a}_{m}(x,\tau) 
  \gamma_{i} \psi^{a}_{m}(x,\tau)] 
  [\overline{\psi}^{b}_{m}(x,\tau^{\prime}) \gamma_{i} 
  \psi^{b}_{m}(x,\tau^{\prime})]  \Biggr ]
\label{quench}
\end{eqnarray}

The loop expansion requires counterterms to account for ultraviolet divergences
in momentum integrals; we write the counter terms as
\begin{eqnarray}
 {\cal S}_{CT} & = & \int d^d x d \tau \Biggl[ \alpha (Z_{\alpha} - 1)
 \overline{\psi}^{a}_{m} \gamma_{0} \partial_{0}
    \psi^{a}_{m} + (Z_2 -1 )\overline{\psi}^{a}_{m} \gamma_{i} 
    \partial_{i} \psi^{a}_{m} \nonumber\\
&&~~~~~~~~~~~~~-\frac{i}{\sqrt{N}} (Z_1^{q} - 1) 
q \mu^{\epsilon/2} \alpha^{1/2} a^{a}_{0}
\overline{\psi}^{a}_{m} \gamma_{0} \psi^{a}_{m}
-\frac{i}{\sqrt{N}} (Z_1^{g} - 1) g \mu^{\epsilon/2} \alpha^{1/2} a^{a}_{i}
\overline{\psi}^{a}_{m} \gamma_{i} \psi^{a}_{m} \Biggr]   
		   \nonumber   \\
  & + & \int d^d x d \tau d \tau^{\prime}
  \Biggl[ ( Z_{M}-1) \Delta_{M} \mu^{\epsilon}
  [\overline{\psi}^{a}_{m}(x,\tau) \psi^{a}_{m}(x,\tau)] 
   [\overline{\psi}^{b}_{m}(x,\tau^{\prime}) \psi^{b}_{m}(x,\tau^{\prime})] 
			 \nonumber   \\
  & &~~~~~~~~~~~~~~~~-(Z_{V}-1) \Delta_{V} \mu^{\epsilon}
  [\overline{\psi}^{a}_{m}(x,\tau) \gamma_{0} \psi^{a}_{m}(x,\tau)] 
  [\overline{\psi}^{b}_{m}(x,\tau^{\prime}) \gamma_{0} 
  \psi^{b}_{m}(x,\tau^{\prime})]   \nonumber   \\
  & &~~~~~~~~~~~~~~~~  +(Z_{A}-1) \Delta_{A} \mu^{\epsilon} 
  [\overline{\psi}^{a}_{m}(x,\tau) 
  \gamma_{i} \psi^{a}_{m}(x,\tau)] 
  [\overline{\psi}^{b}_{m}(x,\tau^{\prime}) \gamma_{i} 
  \psi^{b}_{m}(x,\tau^{\prime})]  \Biggr ]
\label{ct}
\end{eqnarray}

In general, counter terms for the last two gauge field terms in ${\cal S}$ should also be considered.
However, it was shown~\cite{jinwu} that at least to two loops,
there no divergences
associated with these terms. The Ward identities following 
from gauge invariance dictate
\begin{equation}
   Z_1^q = Z_{\alpha},~~~~~ Z_{1}^{g} = Z_2
\label{ward}
\end{equation}

   Using these identities, we relate the bare fields and couplings 
   in ${\cal S}$ to the renormalized quantities by 
\begin{eqnarray}
    \psi_{mB} & = & Z_2^{1/2} \psi_m    \nonumber  \\
    \alpha_B & = &(Z_{\alpha}/Z_2 ) \alpha    \nonumber  \\
     q_B & = & q \mu^{\epsilon/2} (Z_{\alpha} / Z_2 )^{1/2}  \nonumber  \\
     g_B & = & g \mu^{\epsilon/2} (Z_2 / Z_{\alpha} )^{1/2}  \nonumber  \\
     \Delta_{MB} &=&  \mu^{\epsilon} \Delta_{M} Z_{M}/Z^{2}_{2}
			  \nonumber  \\
     \Delta_{VB} &=&  \mu^{\epsilon} \Delta_{V} Z_{V}/Z^{2}_{2}
			  \nonumber  \\
     \Delta_{AB} &=&  \mu^{\epsilon} \Delta_{A} Z_{A}/Z^{2}_{2}
\label{relation}
\end{eqnarray}

Notice that these relations imply that for the statistical angle $\theta/N = qg/N$ we have
$\theta_B = \theta \mu^{\epsilon}$ even in the presence of the
Coulomb interaction
and the disorders; so in $d=2$ this angle is a renormalization group
invariant, which is expected on general physical grounds. 

The dynamic critical exponent, $z$ is related to the renormalization of $\alpha$
by~\cite{cardy}
\begin{equation}
z = 1 - \mu \frac{d}{d\mu} \ln \alpha = 1 - \mu \frac{d}{d\mu} \ln \frac{Z_2}{Z_{\alpha}}
\label{zval}
\end{equation}

We will find it convenient to express the loop expansion in terms of 
the ``fine structure'' constant $w \equiv  q^2 / 16$, and a central object of
study shall be its $\beta$-function $\beta ( w ) = \mu (dw/d\mu)$. By comparing
(\ref{zval}) with relationships between bare and renormalized quantities quoted above
we see that
\begin{equation}
z = 1 - \beta (w)/w.
\label{zbeta}
\end{equation}

Finally, the critical exponent $\nu$ is related to the anomalous dimension
of the composite operator $\overline{\psi} \psi$ by $\nu^{-1} - 1
= \mu (d \ln Z_{\overline{\psi} \psi} / d \mu)$; the renormalization constant
$Z_{\overline{\psi} \psi}$ can be calculated by inserting the 
operator into the self-energy diagrams.

\section{ One-Loop calculation}

We begin the explicit calculation of the renormalization constants by considering
a direct perturbative expansion in the Coulomb fine structure constant $w$,
the statistical angle $\theta$ and the three kinds of disorders 
$ \Delta_{M}, \Delta_{V}, \Delta_{A} $.
At one-loop order (Fig.13, Fig.4a,b,c), we find no dependence
on $\theta$; the values of the renormalization constants upto terms of order
$w^2$, $\theta^2$, $w \theta$ and $ \Delta^{2} $ are
\begin{eqnarray}
Z_2 & = & 1 - 2 w /N \pi \epsilon           \nonumber  \\
Z_{\alpha} & = & 1-\frac{1}{ \pi \epsilon}( \Delta_{M}+ \Delta_{V}
      +2 \Delta_{A} )
\label{oneloopz}
\end{eqnarray}

We also explicitly verified that the Ward identities Eq.\ref{ward} hold. 

  From Fig.5-Fig.10 and Fig.13a-d, we can calculate the three renormalization
  constants for the disorders ( the details are given in the next section where
  we perform a similar calculations in large $ N $ limit ):
\begin{eqnarray}
   Z_{M} & = & 1+ \frac{1}{\pi \epsilon \Delta_{M}}
      (2 \Delta^{2}_{M} +2 \Delta_{M} \Delta_{V}
      -4 \Delta_{M} \Delta_{A}
      -4 \Delta_{V} \Delta_{A}) -\frac{8 w}{N \pi \epsilon}
          \nonumber   \\
   Z_{V} & = & 1- \frac{2}{\pi \epsilon \Delta_{V}}
      ( \Delta_{M} \Delta_{V} +2 \Delta_{M} \Delta_{A}
      + \Delta_{V} \Delta_{V}
      + 2 \Delta_{V} \Delta_{A} )    \nonumber  \\
   Z_{A} & = & 1- \frac{2}{\pi \epsilon \Delta_{A}}
       \Delta_{M} \Delta_{V}
      -\frac{4 w}{ N \pi \epsilon}  
\end{eqnarray}

%~~~~~Z_{\overline{\psi} \psi} = Z_2.
    From Eq.\ref{relation}, we obtain the four $ \beta $ functions:
\begin{eqnarray}
  \beta(w) & = &  \beta^{p}(w)-\frac{w}{\pi}
   ( \Delta_{M}+\Delta_{V}+2 \Delta_{A})    \nonumber  \\
  \beta(\Delta_{M}) & = &-2 \Delta_{M}(\nu^{-1}_{p}-1)+\frac{2}{\pi} (
    \Delta^{2}_{M}+ \Delta_{M}\Delta_{V}- 2\Delta_{M} \Delta_{A}
    -2\Delta_{V}\Delta_{A})    \nonumber  \\
  \beta(\Delta_{V}) & = & 2 \Delta_{V} \frac{\beta^{p}(w)}{w}-\frac{2}{\pi} (
     \Delta_{M} \Delta_{V} + 2 \Delta_{M} \Delta_{A}   
     +\Delta^{2}_{V}+ 2\Delta_{V} \Delta_{A})           \nonumber  \\
  \beta(\Delta_{A}) & = & -\frac{2}{\pi} \Delta_{M} \Delta_{V}
\label{beta1}
\end{eqnarray}

   Where the $\beta$-function of the Coulomb coupling
   in the {\em pure} system  $\beta^{p}(w) $ is\cite{jinwu}:
\begin{equation}
\beta^{p} (w ) = \frac{2 w^2}{N \pi} + {\cal O}( w^3 , w^2 \theta^2 )
\label{oneloopbeta}
\end{equation}

  While the {\em pure} critical exponent $ \nu_{p} $ is \cite{jinwu}:
\begin{equation}
   \nu_{p} = 1 - 2 w /N \pi
\label{oneloopexp}
\end{equation}
 
  From Eq.\ref{beta1}, we can identify the {\em disordered} critical
  exponent $ \nu $:
\begin{equation}
   \nu^{-1}= \nu_{p}^{-1}-\frac{\Delta_{M}}{\pi}-\frac{\Delta_{V}}{\pi}
     + \frac{ 2 \Delta_{A}}{\pi}
\label{dis1}
\end{equation}

    From the first and the third equations in Eq.\ref{beta1}, we find 
    $\beta(\Delta_{V}) $ and $ \beta(w) $ are related by
\begin{equation}
  \beta(\Delta_{V})  =  2 \Delta_{V} \frac{\beta(w)}{w}
  -\frac{4}{\pi} \Delta_{M} \Delta_{A}   
\label{vw}
\end{equation}

    In Eq.\ref{beta1}, setting Coulomb interaction and
    two of the random potentials to be zero,
    we reproduce the results of Ludwig {\sl et.al} \cite{lud}.

  With the presence of Coulomb interaction $ w \neq 0 $,
  we discuss the three cases separately:

\underline{ $ \Delta_{M} \neq 0 $ }

  The system flows to a line of {\em stable} fixed points given by  
  $ \Delta_{M}=\frac{2w}{N} $. The flow trajectory is given by 
  $ \Delta_{M}=\frac{C}{w^{2}} $, $ C $ is an arbitrary constant(Fig.16a).
   From Eqs.\ref{oneloopexp}, \ref{dis1}, we find $ \nu=1 $ along this line,
   it saturates the bound $ \nu \geq 2/d $ ( $ d $ is the special dimension ).

\underline{ $ \Delta_{V} \neq 0 $ }

  There exists a line of {\em unstable} fixed points given by  
  $ \Delta_{V}=\frac{2w}{N} $. The flow trajectory is given by 
  $ \Delta_{V}=Cw^{2} $ (Fig.16b). Below this line, the system flows
  to the origin, above this line, the system flows to strong coupling
  regime. From Eqs.\ref{oneloopexp}, \ref{dis1}, 
  we find $ \nu=1 $ along this line.

\underline {$ \Delta_{A} \neq 0 $ }

  The system flows to a line of {\em stable} fixed points given by  
  $ \Delta_{A}=\frac{w}{N} $. The flow trajectory is given by 
  $ \Delta_{A}=C $ (Fig.16c). 
  From Eqs.\ref{oneloopexp}, \ref{dis1}, we find
  $ \nu=1-\frac{4 \Delta_{A}}{\pi} $ which continuously changes
  along this line. It should be noted
  that $ \nu < 1 $ along this line.

  From Eq.\ref{zbeta}, it
  is easy to see that $ z=1 $ on all these line of fixed points.
It is also worth noting that, despite the value $z=1$, the critical
correlators are {\em not} Lorentz invariant.

  In the generic case, we do not find any perturbative accessible
  {\em stable} fixed point. The system flows to the strong coupling regime.
  In the next section, we resort to the large $ N $ expansion to explore
  the strong coupling regime.

\section{ Large $ N $ expansion }

To understand larger values of $\theta$, and to explore the consequences of a 
possible interference between the Coulomb interactions, disorders
and the Chern-Simons term
we found it convenient to perform a $1/N$ expansion. This is technically
simpler than a perturbative two-loop extension of the computation of the last
section
and also automatically includes the dynamic screening of the gauge field propagator
by the fermion polarization~\cite{bert}. Alternatively stated, the so-called RPA
approximation becomes exact at $N=\infty$, and $1/N$ corrections require
gauge field propagators which have the RPA form
\begin{equation}
{\cal S}_{RPA} = \frac{1}{2} \int \frac{d^2 k}{4 \pi^2} \frac{d \omega}{2 \pi}
(a_0 , a_t)
\left( \begin{array}{cc}
q^2 k^2 /(16 \sqrt{k^2 + \omega^2}) & i k \\
i k & k + g^2 /(16 \sqrt{k^2 + \omega^2}) \end{array}
\right) \left( \begin{array}{c} a_0 \\ a_t \end{array} \right)
\label{lrpa}
\end{equation}

   The inverse of the above matrix leads to the propagators of the
   gauge fields:

\begin{eqnarray}
  G_{00}(q,\nu) &=& \frac{1}{q^{2}} \frac{q + \frac{ g^{2}}{16}
    \sqrt{q^{2}+\nu^{2}}}
      {1+(\frac{\theta}{16})^{2}+
      \frac{e^{2}}{16} \frac{q}{\sqrt{q^{2}+\nu^{2}}}}  \nonumber  \\
  G_{0i}(q,\nu) &=& \frac{\epsilon_{ij} q_{j}}{q^{2}}
     \frac{1} {1+(\frac{\theta}{16})^{2}+
      \frac{e^{2}}{16} \frac{q}{\sqrt{q^{2}+\nu^{2}}}}
                             \nonumber  \\
  G_{0i}(q,\nu) &=& -G_{i0}(q,\nu)   \nonumber   \\
  G_{ij}(q,\nu) &=& (\delta_{ij}-\frac{q_{i} q_{j}}{q^{2}})
     \frac{e^{2}/16} { [1+(\frac{\theta}{16})^{2}]\sqrt{ q^{2}+\nu^{2}}+
      \frac{e^{2}}{16} q}
\label{gauge}
\end{eqnarray}

   To the order of $ (\frac{1}{N})^{0} \Delta $, the summation
   of Fig.1 (a),(b),(c),(d),(e) leads to {\em effective} random scalar
   potential $ \tilde{\Delta}_{V} $:

\begin{equation} 
  \tilde{\Delta}_{V} = \frac{ \Delta_{V}  + \Delta_{A} \phi}
  { (\lambda+ w)^{2} } 
\end{equation}
where $ \phi \equiv (\theta/16)^{2}$ and $\lambda=1+ \phi $. 

    Note that both $ \Delta_{V} $ and $\Delta_{A} $
    contribute to the effective random scalar potential with the {\em same}
    sign.

   To the order of $ (\frac{1}{N})^{0} \Delta $, the summation
   of Fig.2 (a),(b),(c),(d),(e) leads to {\em effective} random vector
   potential
   $ \tilde{\Delta}_{A} (\delta_{ij}-\frac{q_{i} q_{j} }{q^{2}}) $
   (see the appendix):

\begin{equation} 
  \tilde{\Delta}_{A} = \frac{ \Delta_{V} \phi +
     \Delta_{A} (1+w)^{2}}{ (\lambda+ w)^{2} } 
\end{equation}

    Note that both $ \Delta_{A} $ and $\Delta_{V} $
    contribute to the effective random vector potential
    with the {\em same} sign.

   To the order of $ (\frac{1}{N})^{0} \Delta $, the summation
   of Fig.3 (a),(b),(c),(d) leads to {\em effective} random Scalar-Vector
   potential $ \tilde{\Delta}_{C} \frac{\epsilon_{ij} q_{j}}{q} $:

\begin{equation} 
  \tilde{\Delta}_{C} = \frac{2 \sqrt{\phi} } { (\lambda+ w)^{2} } [
  -\Delta_{V} + \Delta_{A}  (1+ w) ]
\end{equation}

   It should be noted that both $ \Delta_{A} $ and $\Delta_{V} $
   contribute to the effective random SV potential
   with the {\em opposite} sign and
   there is {\em no} such random SV
   potential in the original action Eq.\ref{quench}. Due to the interference
   between disorder and CS interaction, this potential is generated. But
   it can only appear as {\em internal} lines.
   Putting $ \phi=0 $ (namely, no CS interaction), this potential vanishes.

   These three effective potentials plus the random mass potential and
   the four RPA gauge fields propagators in Eq.\ref{gauge} are the building
   blocks in the following Feymann diagrams. Note that, however, only {\em bare}
   disorder potentials can appear as {\em external} lines of any Feymann diagram.

      In the following, we discuss $ N=\infty $ case first, then we
      discuss $ 1/N $ correction. It is well known that only {\em primitive}
      divergences of Feymann diagrams are needed in the RG flow equation.

\subsection{ $ N=\infty $ limit}

     Fig.4 is the contributions to the Dirac fermion self-energy
     from the effective random potentials. The divergent parts 
     from the effective random SV potentials (d) and (e) vanish. We find
     the two constants defined in Eq.\ref{ct}:
\begin{eqnarray}
  Z_{2} & = & 1     \nonumber  \\
  Z_{\alpha} & = & 1-\frac{1}{ \pi \epsilon}( \Delta_{M}+ \tilde{\Delta}_{V}
      +2 \tilde{\Delta}_{A} )
\end{eqnarray}

  In the following, the notation $ A \rightarrow B $ means that $ A $
  renormalizes $ B $.

 Fig.5 is the renormalization from the random mass ( In this figure and the
 following figures, we do not draw explicitly the diagrams with the
 exchange of leg3 and leg 4). 

\begin{eqnarray}
 & 5a &  = -5b = \frac{\Delta_{M}^{2}}{\pi \epsilon} 
     \rightarrow \Delta_{A}        \nonumber  \\
  & 5c & = 5d = \frac{2 \Delta_{M}^{2}}{\pi \epsilon} 
  \rightarrow \Delta_{M} 
\end{eqnarray}
   In all, Fig.5 contributes $ \frac{ 4 \Delta_{M}^{2}}{\pi \epsilon}
  \rightarrow \Delta_{M} $.

 Fig.6 is the renormalization from the interference of random mass 
 and {\em effective} random scalar potential. Note the effective
 random scalar potential can only appear in the {\em internal loops}.

\begin{eqnarray}
  & 6a & = 6b = 6c = 6d =  -\frac{\Delta_{M} \tilde{\Delta}_{V}}
   {\pi \epsilon} \rightarrow \Delta_{A}    \nonumber   \\
  & 6e & = 6f = \frac{2 \Delta_{M} \tilde{\Delta}_{V}}
   {\pi \epsilon} \rightarrow \Delta_{M}   \nonumber  \\
  & 6g & = 6h = \frac{2 \Delta_{M} \Delta_{V}}
   {\pi \epsilon} \rightarrow \Delta_{V}     
\end{eqnarray}
   In all, Fig.6 contributes $\frac{4 \Delta_{M} \tilde{\Delta}_{V}}
   {\pi \epsilon} \rightarrow \Delta_{M}, 
   \frac{4 \Delta_{M} \Delta_{V}} {\pi \epsilon} \rightarrow \Delta_{V},
   -\frac{4 \Delta_{M} \tilde{\Delta}_{V}}
   {\pi \epsilon} \rightarrow \Delta_{A} $

 Fig.7 is the renormalization from the interference of random mass 
 and {\em effective} random vector potential.
\begin{eqnarray}
  & 7a & = 7b = \frac{2 \Delta_{M} \tilde{\Delta}_{A}}
   {\pi \epsilon} \rightarrow \Delta_{M},
   \frac{2 \Delta_{M} \tilde{\Delta}_{A}}
   {\pi \epsilon} \rightarrow \Delta_{V}    \nonumber  \\
  & 7c & = 7d = -\frac{2 \Delta_{M} \tilde{\Delta}_{A}}
   {\pi \epsilon} \rightarrow \Delta_{M},
   \frac{2 \Delta_{M} \tilde{\Delta}_{A}}
   {\pi \epsilon} \rightarrow \Delta_{V}   \nonumber  \\
  & 7e & = 7f = - \frac{ 4 \Delta_{M} \tilde{\Delta}_{A}}
   {\pi \epsilon} \rightarrow \Delta_{M}  \nonumber  \\
   & 7g & = 7h = 0
\end{eqnarray}
   
   In all, Fig.7 contributes
    $ -\frac{8 \Delta_{M} \tilde{\Delta}_{A}}
   {\pi \epsilon} \rightarrow \Delta_{M} $, 
    $  \frac{8 \Delta_{M} \tilde{\Delta}_{A}}
   {\pi \epsilon} \rightarrow \Delta_{V} $. 

 Fig.8 is the renormalization from the random scalar potential. 
\begin{eqnarray}
  & 8a & = -8b = \frac{\tilde{\Delta}^{2}_{V}}{\pi \epsilon} 
  \rightarrow \Delta_{A}   \nonumber  \\
  & 8c & = 8d= \frac{2 \Delta_{V} \tilde{\Delta}_{V}}
   {\pi \epsilon} \rightarrow \Delta_{V}  
\end{eqnarray}

    In all, Fig.8 contributes 
   $ \frac{ 4 \Delta_{V} \tilde{\Delta}_{V}}
   {\pi \epsilon} \rightarrow \Delta_{V} $. 

 Fig.9 is the renormalization from the interference of effective random
 scalar potential and effective random vector potential.
\begin{eqnarray}
  & 9a & = 9b = -\frac{2 \tilde{\Delta}_{V} \tilde{\Delta}_{A}}
   {\pi \epsilon} \rightarrow \Delta_{M},
    -\frac{2 \tilde{\Delta}_{V} \tilde{\Delta}_{A}}
   {\pi \epsilon} \rightarrow \Delta_{V}  \nonumber  \\
  & 9c & = 9d = -\frac{2 \tilde{\Delta}_{V} \tilde{\Delta}_{A}}
   {\pi \epsilon} \rightarrow \Delta_{M},
    \frac{2 \tilde{\Delta}_{V} \tilde{\Delta}_{A}}
   {\pi \epsilon} \rightarrow \Delta_{V}  \nonumber  \\
  & 9e & = 9f = \frac{4 \Delta_{V} \tilde{\Delta}_{A}}
   {\pi \epsilon} \rightarrow \Delta_{V}    \nonumber  \\
  & 9g & = 9h=0
\end{eqnarray}
    In all, Fig.9 contributes 
   $- \frac{ 8 \tilde{\Delta}_{V} \tilde{\Delta}_{A}}
   {\pi \epsilon} \rightarrow \Delta_{M},
   \frac{ 8 \Delta_{V} \tilde{\Delta}_{A}}
   {\pi \epsilon} \rightarrow \Delta_{V} $.

 Fig.10 is the renormalization from the effective random vector potential. 
\begin{eqnarray}
  & 10a & = - 10b =  \frac{4 \tilde{\Delta}^{2}_{A}}{\pi \epsilon} 
   \rightarrow \Delta_{A}            \nonumber  \\
   & 10c & = 10d = 0
\end{eqnarray}

    In all, Fig.10 does {\em not} contribute. 

 Fig.11 is the renormalization from the effective random SV potential. 
\begin{eqnarray}
 & 11a & = 11d = - 11e = -11h =
   -\frac{\tilde{\Delta}^{2}_{C}}{ 4 \pi \epsilon} \rightarrow \Delta_{A}
			   \nonumber   \\
  &11b & = 11c = 11f = 11g=
  -\frac{\tilde{\Delta}^{2}_{C}}{ 2 \pi \epsilon} \rightarrow \Delta_{M} 
\end{eqnarray}
    In all, Fig.11 contributes 
   $- \frac{ 2 \tilde{\Delta}^{2}_{C} }{\pi \epsilon} \rightarrow \Delta_{M} $.
   It can be shown that only when $ SV $ potential get paired, there are
   {\em non-zero } contributions.

   Fig.12 is the renormalization to the random mass from random effective
   scalar, vector and scalar-vector potentials.
\begin{eqnarray}
  &  12a &= 12h= \frac{ 16 \Delta_{M} \tilde{\Delta}_{A} }{ \pi \epsilon}
    \frac{w+\phi}{\lambda+w}  \rightarrow \Delta_{M}      \nonumber  \\
  &  12b &= 12c= 12f =12g = \frac{8 \Delta_{M} \tilde{\Delta}_{C} }
  { \pi \epsilon} \frac{\sqrt{\phi}}{\lambda+w}  
		   \rightarrow   \Delta_{M} \nonumber  \\
  &  12d &= 12e= - \frac{ 16 \Delta_{M} \tilde{\Delta}_{V} }{ \pi \epsilon}
    \frac{\phi}{\lambda+w}  \rightarrow \Delta_{M}
\end{eqnarray}
    In all, Fig.12 contributes $   - \frac{32 \Delta_{M}}{\pi \epsilon}
    \frac{1}{ \lambda+w} [ \phi \tilde{\Delta}_{V}
    -(w+\phi) \tilde{\Delta}_{A}
    -\sqrt{\phi} \tilde{\Delta}_{C} ]   \rightarrow  \Delta_{M} $.

  From Fig5-Fig.12, we find the three constants $ Z_{M}, Z_{V}, Z_{A} $
  defined in Eq.\ref{ct}:
\begin{eqnarray}
   Z_{M} & = & 1+ \frac{1}{\pi \epsilon \Delta_{M}}
      (2 \Delta^{2}_{M} +2 \Delta_{M} \tilde{\Delta}_{V}
      -4 \Delta_{M} \tilde{\Delta}_{A}
      -4 \tilde{\Delta}_{V} \tilde{\Delta}_{A}
      - \tilde{\Delta}^{2}_{C}-16 \Delta_{M} \tilde{\Delta}_{K} ) 
      \nonumber   \\
   Z_{V} & = & 1- \frac{2}{\pi \epsilon \Delta_{V}}
      ( \Delta_{M} \Delta_{V} +2 \Delta_{M} \tilde{\Delta}_{A}
      + \Delta_{V} \tilde{\Delta}_{V}
      + 2 \Delta_{V} \tilde{\Delta}_{A} )   \nonumber  \\
   Z_{A} & = & 1- \frac{2}{\pi \epsilon \Delta_{A}}
       \Delta_{M} \tilde{\Delta}_{V}
\end{eqnarray}
     Where $ \tilde{\Delta}_{K} $ is given by
\begin{equation} 
  \tilde{\Delta}_{K} = \frac{ \Delta_{V} \phi (3-w-\phi) +
     \Delta_{A} [\phi^{2}-(w+\phi) (1+w)^{2}-2 \phi (1+w)]}{ (\lambda+ w)^{3} } 
\end{equation}

  From Eq.\ref{relation}, we can find the following $ \beta $ functions
  at $ N= \infty $.
\begin{eqnarray}
  \beta(w) &=& -\frac{w}{\pi}( \Delta_{M}+ P(\tilde{\Delta}_{V})
	       + 2 P( \tilde{\Delta}_{A}) )   \nonumber  \\
  \beta(\Delta_{M}) &=&\frac{2 }{\pi} \Delta^{2}_{M} 
     +\frac{1 }{\pi} \tilde{\Delta}^{2}_{C} 
     +\frac{4 }{\pi} \tilde{\Delta}_{V} \tilde{\Delta}_{A} 
     +\frac{2 }{\pi} (\Delta_{M} - 2\tilde{\Delta}_{A} )
     P( \tilde{\Delta}_{V} ) 
     -\frac{4 }{\pi} (\Delta_{M} + \tilde{\Delta}_{V} )
     P( \tilde{\Delta}_{A} )    \nonumber   \\
    & - &\frac{2 }{\pi} \tilde{\Delta}_{C} P( \tilde{\Delta}_{C} ) 
     -\frac{16}{\pi} \Delta_{M} P(\tilde{\Delta}_{K})  \nonumber  \\
  \beta(\Delta_{V}) &=& -\frac{2 }{\pi} \Delta_{M} \Delta_{V} 
     -\frac{2 }{\pi} \Delta_{V} P( \tilde{\Delta}_{V} ) 
     -\frac{4 }{\pi} (\Delta_{M} + \Delta_{V} ) P( \tilde{\Delta}_{A} ) 
					\nonumber   \\
  \beta(\Delta_{A}) &=& -\frac{2 }{\pi} \Delta_{M} P( \tilde{\Delta}_{V} ) 
\label{beta2}
\end{eqnarray}
      Where the function $ P(\Delta)
      =\Delta_{V} \frac{\partial \Delta}{\partial \Delta_{V}}
      +\Delta_{A} \frac{\partial \Delta}{\partial \Delta_{A}}
      + w \frac{\partial \Delta}{\partial w}
      + 2\phi \frac{\partial \Delta}{\partial \phi} $.

    We shall discuss the implications of this equation after
    considering $ 1/N $ corrections in the next subsection.

\subsection{ $ 1/N $ correction }

    In this section, we consider $ 1/N $ correction to $ N=\infty $
    results. The small parameters are $ 1/N, \Delta_{M}, \Delta_{V},
    \Delta_{A} $. We expect that if there are fixed points, 
    the fixed points
    values of $ \Delta^{\prime s} $ are of the order $ 1/N $.

 Fig.13 is the contribution to the Dirac fermion self energy from $ 1/N $
 fluctuation of gauge fields given by Eq.\ref{gauge}. Actually,
 Fig.13b and Fig.13c are convergent. The results are \cite{jinwu}
\begin{eqnarray}
 Z_{2} & = & 1- \frac{\Psi_{A}}{N \pi \epsilon} =
 1-\frac{1}{ N \pi \epsilon} \left( \frac{2 w}{ \lambda}
    -\frac{ 16 w^2 A}{\pi \lambda} + \frac{ \theta^{2} C}{ 16 \pi}
    -\frac{\theta^{2} E}{16 \pi} \right)    \nonumber   \\
 Z_{\alpha} & = & 1-\frac{\Psi_{V}}{N \pi \epsilon} =
    1-\frac{1}{ N \pi \epsilon} \left(
    \frac{16 w^2 B}{\pi \lambda} - \frac{ \theta^{2} D}{ 16 \pi}
    +\frac{\theta^{2} F}{16 \pi} \right),
\label{diverge}
\end{eqnarray}
where $ \lambda=1+ (\theta/16)^{2}$ and the functions $ A,B,C,D, E=A+B, F=B $
are given by the formal expressions
\begin{eqnarray}
  A & = & \int^{1}_{0} d x \frac{ 4 x^{2} (1-x^{2}) }{ (1+x^{2})^{3} }
  f(x;w,\theta),  
  ~~~~~ B= \int^{1}_{0} d x \frac{ (1-x^{2})(1-6 x^{2}+x^{4}) }{ (1+x^{2})^{3} }
  f(x;w,\theta)     \nonumber \\
  C & = & \int^{1}_{0} d x \frac{ 4 x^{2} }{(1-x^{2})
  (1+x^{2}) } f(x;w,\theta),  
  ~~~~~ D= \int^{1}_{0} d x \frac{ (1-6 x^{2}+x^{4}) }
  { (1-x^{2}) (1+x^{2}) } f(x;w,\theta) 
\label{constant}
\end{eqnarray}
with $ f(x;w,\theta)=( \lambda (1+x^{2})+ w(1-x^{2}))^{-1} $, and the variable $x$
represents an intermediate frequency.
Note the two constants $ C, D $ are divergent: this 
divergence is due to the singular effect of frequencies $|\omega| \gg k$. 
However, as shown in Ref~\cite{jinwu}, these divergences are gauge artifacts
and cancel in the $\beta$-function and in
  any physical gauge-invariant quantity like $ \nu$, $z $
  or $ \sigma_{ij}$. The divergences however do infect
the anomalous dimension of the field operator $\psi$:
this is as expected as the propagator of $\psi$ is clearly
gauge-dependent.

  Fig.14 is the renormalization to $ \Delta_{M} $ from the $ 1/N $
 fluctuation of gauge fields. Actually, Fig.14b and Fig.14c are convergent,
 the divergent parts are

\begin{equation}
   \frac{\Psi_{M}}{N \pi \epsilon}= \frac{1}{ N \pi \epsilon} [
   \frac{4 w}{\lambda} (1- \frac{4 w (2A+B) }{\pi} )+
   \frac{\theta^{2} (2 C +D) }{16 \pi} + \frac{ \theta^{2} (2A+B) }{16 \pi} - 
    \frac{ \theta^{2} G}{2 \pi} ]
\end{equation}
    Where the function $ G $ is given by
\begin{equation}
  G  =  \int^{1}_{0} d x [ (\phi-1)(1-x^{2})+ w \frac{ (1-x^{2})^{2} }
       { 1+x^{2} } ] f^{2}(x;w,\theta)  
\end{equation}
    
      It is easy to see  Fig.14 is exactly {\em the same}
      diagram for the calculation of $ Z_{\bar{\psi}\psi} $ as expected
      \cite{jinwu},
      therefore $ Z_{2} Z_{\bar{\psi} \psi}=1-\Psi_{M}/N \pi \epsilon  $.

   In Fig14, replacing $\Delta_{M} $ line by
   $ \Delta_{V} $ and $ \Delta_{A} $ lines,
   we can repeat the same calculation.
   In fact, the divergent parts
   should be $ \frac{\Psi_{V}}{N \pi \epsilon} $ and 
   $ \frac{\Psi_{A}}{N \pi \epsilon} $ respectively as dictated by
   Ward identities.
   It can be shown explicitly that Fig.15a + Fig15b vanishes as
   dictated by Ward Identities.

 Adding the $ 1/N $ correction to the renormalization constants calculated
 in the last section, we obtain

\begin{eqnarray}
  Z_{2} & = & 1-\frac{\Psi_{A}}{N \pi \epsilon}     \nonumber  \\
  Z_{\alpha} & = & 1-\frac{1}{ \pi \epsilon}( \Delta_{M}+ \tilde{\Delta}_{V}
      +2 \tilde{\Delta}_{A}  ) -\frac{\Psi_{V}}{N \pi \epsilon}                                                        \nonumber   \\
   Z_{M} & = & 1+ \frac{1}{\pi \epsilon \Delta_{M}}
      (2 \Delta^{2}_{M} +2 \Delta_{M} \tilde{\Delta}_{V}
      -4 \Delta_{M} \tilde{\Delta}_{A}
      -4 \tilde{\Delta}_{V} \tilde{\Delta}_{A}
      - \tilde{\Delta}^{2}_{C} 
      -16 \Delta_{M} \tilde{\Delta}_{K}  ) -\frac{2 \Psi_{M}}{N \pi \epsilon}
          \nonumber   \\
   Z_{V} & = & 1- \frac{2}{\pi \epsilon \Delta_{V}}
      ( \Delta_{M} \Delta_{V} +2 \Delta_{M} \tilde{\Delta}_{A}
      + \Delta_{V} \tilde{\Delta}_{V}
      + 2 \Delta_{V} \tilde{\Delta}_{A} ) 
      -\frac{2 \Psi_{V}}{ N \pi \epsilon}   \nonumber  \\
   Z_{A} & = & 1- \frac{2}{\pi \epsilon \Delta_{A}}
       \Delta_{M} \tilde{\Delta}_{V}
      -\frac{2 \Psi_{A}}{ N \pi \epsilon}  
\end{eqnarray}

  From Eq.\ref{relation}, we can find $ 1/N $ corrections
   to the $ \beta $ functions in Eq.\ref{beta2}.
\begin{eqnarray}
  \beta(w) &=& \beta^{p}(w)-\frac{w}{\pi}( \Delta_{M}+ P(\tilde{\Delta}_{V})
	       + 2 P( \tilde{\Delta}_{A}) )   \nonumber  \\
  \beta(\Delta_{M}) &=& -2 \Delta_{M}(\nu^{-1}_{p}-1)
     +\frac{2 }{\pi} \Delta^{2}_{M} 
     +\frac{1 }{\pi} \tilde{\Delta}^{2}_{C}  
     +\frac{4 }{\pi} \tilde{\Delta}_{V} \tilde{\Delta}_{A}  
     +  \frac{2 }{\pi} (\Delta_{M} - 2\tilde{\Delta}_{A} )
     P( \tilde{\Delta}_{V} )   \nonumber  \\
    & - & \frac{4 }{\pi} (\Delta_{M} + \tilde{\Delta}_{V} )
     P( \tilde{\Delta}_{A} ) 
     -\frac{2 }{\pi} \tilde{\Delta}_{C} P( \tilde{\Delta}_{C} ) 
     -\frac{16}{\pi} \Delta_{M} P(\tilde{\Delta}_{K})    \nonumber  \\
  \beta(\Delta_{V}) &=& 2 \Delta_{V} \frac{\beta^{p}(w)}{w}
     -\frac{2 }{\pi} \Delta_{M} \Delta_{V} 
     -\frac{2 }{\pi} \Delta_{V} P( \tilde{\Delta}_{V} ) 
     -\frac{4 }{\pi} (\Delta_{M} + \Delta_{V} ) P( \tilde{\Delta}_{A} ) 
					\nonumber   \\
  \beta(\Delta_{A}) &=& -\frac{2 }{\pi} \Delta_{M} P( \tilde{\Delta}_{V} ) 
\label{beta3}
\end{eqnarray}
      
     The $ \beta^{p}(w) $ in  Eq.\ref{beta3}
  is the $\beta$-function of the Coulomb coupling in the {\em pure} case
  which was calculated in Ref.\cite{jinwu}:
\begin{eqnarray}
&& \beta^{p} (w) = \frac{2 w^{2} (1-\phi)}{ N \pi^{2} 
   \lambda^{2}} \left[ \pi-16 w \int^{1}_{0} d x \left(\frac{ 1-x^{2}}{1+x^{2}}
\right)^{3}
   \frac{ \lambda (1+x^{2})+ \frac{ w}{2}(1-x^{2}) }{( \lambda (1+x^{2})
   +w (1-x^{2}))^{2}} \right]  \nonumber   \\
    &&~~~~~~ +  \frac{32 w \phi}{ N \pi^{2}}
     \int^{1}_{0} d x \frac{( 1-x^{2})(-1+10 x^{2}-x^{4} )}{(1+x^{2})^{3}}
   \frac{ (1+x^{2})+ \frac{ w}{2}(1-x^{2}) }{( \lambda (1+x^{2})
   +w (1-x^{2}))^{2}},
\label{complete}
\end{eqnarray}

 The {\em pure} exponent $\nu_{p} $ in Eq.\ref{beta3} is given by
\begin{eqnarray}
 \nu^{-1}_{p}-1 & = & \frac{2 w (1-\phi)}{ N \pi^{2} \lambda^{2}} 
 \left[ \pi-16 w \int^{1}_{0} d x \left(\frac{ 1-x^{2}}{1+x^{2}} \right)^{3}
   \frac{ \lambda (1+x^{2})+ \frac{ w}{2}(1-x^{2}) }{( \lambda (1+x^{2})
   +w (1-x^{2}))^{2}} \right]  \nonumber   \\
 & + & \frac{32 \phi}{ N \pi^{2} } \int^{1}_{0} d x 
 (\frac{ 1-x^{2}}{1+x^{2}})^{3}
   \frac{ 1+x^{2}+ \frac{ w}{2}(1-x^{2}) }{( \lambda (1+x^{2})
   +w (1-x^{2}))^{2}}  \nonumber   \\
 &- &\frac{192 \phi}{ N \pi^{2} } \int^{1}_{0} d x (\frac{ 1-x^{2}}{1+x^{2}})
   \frac{ 1+x^{2}+ \frac{ w}{2}(1-x^{2}) }{( \lambda (1+x^{2})
   +w (1-x^{2}))^{2}}  \nonumber   \\
 & + & \frac{512 \phi (1- \phi)}{ N \pi^{2}}
     \int^{1}_{0} d x \frac{( 1-x^{2})(1+ x^{2})}
   {( \lambda (1+x^{2}) +w (1-x^{2}))^{3}}
\label{kick}
\end{eqnarray}

    The explicit expressions for $ P^{\prime s} $ are given by
\begin{eqnarray}
   P( \tilde{\Delta}_{V} ) & = & \frac{1}{ (\lambda+w)^{3} }
    [ \Delta_{V}(1-w-3\phi)+\Delta_{A} \phi(3+w-\phi) ] \nonumber  \\
   P( \tilde{\Delta}_{A} ) & = & \frac{1}{ (\lambda+w)^{3} }
    [ \Delta_{V} \phi(3+w-\phi) +\Delta_{A}(1+w)[(1+w)^{2}
     -(3+w)\phi]] \nonumber  \\
   P( \tilde{\Delta}_{C} ) & = & \frac{2 \sqrt{\phi}}{ (\lambda+w)^{3} }
   [- 2 \Delta_{V} (1-\phi) +\Delta_{A} [(\lambda+w)w+2(1+w)(1-\phi)]]
		\nonumber   \\
   P( \tilde{\Delta}_{K} ) & = & \frac{1}{(\lambda+w)^{3} }
   \{ \Delta_{V} \phi (9-4w-5\phi) +\Delta_{A} [5 \phi^{2}
   -(1+w)^{2}(2w+3\phi) 
   -  6 \phi (1+w) -2 w (w+\phi)(1+w)  -  2 \phi w] \} 
		     \nonumber  \\
   & - & \frac{3(w+2 \phi)}{(\lambda+w)^{4}} \{ \Delta_{V} \phi
	(3-w-\phi)+\Delta_{A} [ \phi^{2}-(w+\phi)(1+w)^{2}-2 \phi (1+w)]\}
\label{pfun}
\end{eqnarray}

  From the second equation in Eq.\ref{beta3},
  we can identify the {\em disordered} critical exponent
\begin{equation}
   \nu^{-1}=\nu^{-1}_{p}-\frac{\Delta_{M}}{\pi}-
   \frac{P(\tilde{\Delta}_{V})}{\pi}
   + \frac{ 2 P(\tilde{\Delta}_{A})}{\pi} 
   + \frac{ 8 P(\tilde{\Delta}_{K})}{\pi} 
\label{dis2}
\end{equation}

  At fixed points, substituting $ \frac{\beta^{p}(w)}{w} =\frac{1}{\pi}
  ( \Delta_{M}+ P(\tilde{\Delta}_{V})
       + 2 P( \tilde{\Delta}_{A}) ) $ into Eq.\ref{kick}, we can
       simplify the above equation to
\begin{equation}
   \nu^{-1}= \tilde{\nu}^{-1}_{p}
   + \frac{ 4 P(\tilde{\Delta}_{A})}{\pi} 
   + \frac{ 8 P(\tilde{\Delta}_{K})}{\pi} 
\label{point}
\end{equation}
    Where $ \tilde{\nu}^{-1}_{p} $ is
     listed in Eq.12 of Ref.\cite{jinwu}:
\begin{eqnarray}
 \tilde{\nu}^{-1}_{p}-1 & = & 
  -\frac{128 \phi}{ N \pi^{2} } \int^{1}_{0} d x 
 \frac{ (1-x^{2})(1+6 x^{2} +x^{4})}{(1+x^{2})^{3}}
   \frac{ 1+x^{2}+ \frac{ w}{2}(1-x^{2}) }{( \lambda (1+x^{2})
   +w (1-x^{2}))^{2}}  \nonumber   \\
 & + & \frac{512 \phi (1- \phi)}{ N \pi^{2}}
     \int^{1}_{0} d x \frac{( 1-x^{2})(1+ x^{2})}
   {( \lambda (1+x^{2}) +w (1-x^{2}))^{3}}
\label{kick1}
\end{eqnarray}

   It can be checked that Eq.\ref{vw} should be replaced by:
\begin{equation}
  \beta(\Delta_{V})  =  2 \Delta_{V} \frac{\beta(w)}{w}
  -\frac{4}{\pi} \Delta_{M} P(\tilde{\Delta}_{A})   
\label{vw1}
\end{equation}

  In Eq.\ref{beta3}, we expand $ \beta(w) $ to order $ 1/N $ and $ \Delta $,
  expand $ \beta(\Delta) $ to $ ( 1/N ) \Delta, \Delta^{2} $. It is easy to see
  there should be {\em no} $ (1/N)^{2} $ terms in $ \beta(\Delta) $,
  because interactions do not generate disorder (or equivalently disorder
  do not generate interactions ).
  Note that the small parameter $ 1/N $ plays a similar role to the
  small parameter $ \epsilon $ in the conventional $ \epsilon $ expansion,
  namely, we are trying to locate the fixed points at $ \Delta^{\prime s} $
  at the order of $ 1/N $.

We now turn to the physical implications of our main
results (\ref{beta3} ), (\ref{complete}) and (\ref{kick}).

    If there exists only random mass, namely,
    $ \Delta_{V}=\Delta_{A}=0 $, Eq.\ref{beta3} simplifies to
\begin{eqnarray}
  \beta(w) &=& \beta^{p}(w)-\frac{w}{\pi} \Delta_{M}   \nonumber  \\
  \beta(\Delta_{M}) &=& -2 \Delta_{M}(\nu^{-1}_{p}-1)
     +\frac{2 }{\pi} \Delta^{2}_{M} 
   = -2 \Delta_{M}(\tilde{\nu}^{-1}_{p}-1)
\end{eqnarray}

  Setting all the disorders vanishing, the authors in Ref.\cite{jinwu} found
  a line of {\em pure} fixed points given by  $ \beta^{p}(w)=0 $. They also
  found that $ \tilde{\nu}_{p} > 1 $ along the fixed line,
  therefore concluded that this line is stable against weak random mass disorder
     from Harris criterion. Here, we explicitly write down
     $ \beta(\Delta_{M}) $ to the order $ (1/N)\Delta_{M},
     \Delta^{2}_{M} $ and reach a {\em stronger} statement that
     there are {\em no} other fixed points except this line in the
     weak coupling regime.

  Comparing Eq.\ref{beta2} to Eq.\ref{beta3},
  we find that there is {\em no} $1/N $ correction to $ \beta(\Delta_{A}) $,
  therefore $ \Delta_{A} $ is always marginal.
  The $ 1/N $ correction to $ \beta(\Delta_{V}) $ is simply
  $ \beta^{p}(w)/w $, therefore $ \Delta_{V} $ is marginal along this line,
  irrelevant(relevant) {\em above(below)} this line.
  Actually, these results are expected from Ward identities.
  We also find that $ 1/N $ correction to $ \beta(\Delta_{M}) $ is
  just $ \nu^{-1}_{p}-1 $, which is consistent with Harris criterion.

   With the presence of all the three disorders, we expect that
   the {\em pure} fixed line is unstable.

\subsubsection{Integer Quantum Hall Transition ($ \phi=0 $) }

First, considering the transition out of the integer quantum Hall state, $\theta = 0$,
which implies $\phi=0$, $\lambda = 1$. 
The $ \beta^{p}(w) $ is simplified to \cite{jinwu}
\begin{equation}
 \frac{ \beta^{p} (w)}{w} = \frac{2 w} { N \pi^{2} } 
 \left[ \pi-16 w \int^{1}_{0} d x \left(\frac{ 1-x^{2}}{1+x^{2}}
\right)^{3}
   \frac{  1+x^{2}+ \frac{ w}{2}(1-x^{2}) }{( 1+x^{2}
   +w (1-x^{2}))^{2}} \right]
\label{simple}
\end{equation}

 Eq.\ref{kick} reduces to $ \nu_{p}^{-1}-1= \beta^{p}(w)/w $.
The simple analysis of (\ref{complete})
shows that $\beta^{p} (w) > 0$ for all $w > 0$; 
for small $w$ we have $\beta^{p} (w)/w =
2 w / (N \pi)$, in agreement with one-loop result (\ref{oneloopbeta}),
while for $w \gg 1$,
$\beta (w) = 4 / (N \pi w)$. So the only {\em pure} fixed point
remains at $w = 0$. The whole picture of $ \beta^{p}(w)/w $ is
drawn in Fig.17a,c, it increases linearly first, reaches a maximum value
  $ 0.20/N $ at $ w=1.31 $ and
 eventually decays as $ 1/w $.

  It is easy to see that $\tilde{\Delta}_{V}=\frac{\Delta_{V}}{ (1+w)^{2} },
  P( \tilde{\Delta}_{V} ) = \Delta_{V} \frac{ 1-w }{ (1+w)^{3} };
  \tilde{\Delta}_{A}= P( \tilde{\Delta}_{A} ) = \Delta_{A};
  \tilde{\Delta}_{C}= P( \tilde{\Delta}_{C} ) = 0;
  \tilde{\Delta}_{K}= -\Delta_{A} \frac{w}{ 1+w },
  P( \tilde{\Delta}_{K} ) = -\Delta_{A} \frac{ w(w+2) }{ (1+w)^{2} } $.
  Substituting these expressions into Eq.\ref{beta3}. we find that
  Eq.\ref{beta3} is simplified to 

\begin{eqnarray}
 \frac{ \beta(w)}{w} &=& \frac{\beta^{p}(w)}{w}-
 \frac{1}{\pi}( \Delta_{M}+ \Delta_{V} \frac{1-w}{(1+w)^{3}}
	       + 2 \Delta_{A} )   \nonumber  \\
  \beta(\Delta_{M}) &=& -2 \Delta_{M} \frac{\beta^{p}(w)}{w}
     +\frac{2 }{\pi} \Delta^{2}_{M} 
     +  \frac{2 }{\pi} (\Delta_{M}-2 \Delta_{A})
     \Delta_{V} \frac{1-w}{(1+w)^{3}} 
      + \frac{4 }{\pi} \Delta_{M} \Delta_{A} (3-\frac{4}{(1+w)^{2}})
        \nonumber  \\
  \beta(\Delta_{V}) &=& 2 \Delta_{V} \frac{\beta^{p}(w)}{w}
     -\frac{2 }{\pi} \Delta_{M} \Delta_{V} 
     -\frac{2 }{\pi} \Delta^{2}_{V} \frac{1-w}{(1+w)^{3}}
     -\frac{4 }{\pi} (\Delta_{M} + \Delta_{V} ) \Delta_{A}  
					\nonumber   \\
  \beta(\Delta_{A}) &=& -\frac{2 }{\pi} \Delta_{M} \Delta_{V}
	    \frac{1-w}{(1+w)^{3}}
\label{betac}
\end{eqnarray}

   Eq.\ref{dis2} is simplified to
\begin{equation}
   \nu^{-1}=\nu^{-1}_{p}-\frac{\Delta_{M}}{\pi}-
   \frac{\Delta_{V}}{\pi} \frac{1-w}{(1+w)^{3}}
   + \frac{ 2 \Delta_{A}}{\pi} 
   - \frac{ 8 \Delta_{A}}{\pi} \frac{ w(w+2)}{(1+w)^{2}} 
\label{disc}
\end{equation}

    Eq.\ref{vw1} is simplified to
\begin{equation}
  \beta(\Delta_{V})  =  2 \Delta_{V} \frac{\beta(w)}{w}
  -\frac{4}{\pi} \Delta_{M} \Delta_{A}   
\label{vwc}
\end{equation}

 We discuss the three cases separately:

\underline{ $ \Delta_{M} \neq 0 $ }

  The system flows to a line of {\em stable} fixed points given by  
  $ \Delta_{M}= \pi \frac{\beta^{p}(w)}{w} $. Like the one-loop result,
  $ \nu=1 $ and the flow trajectory is given by 
  $ \Delta_{M}=\frac{C}{w^{2}} $, $ C $ is an arbitrary constant
  $ \sim 1/N $ (Fig.17a). We suspect that $ \nu=1 $ is {\em exact}
  (namely, independent of large $ N $ limit ).

   This line is {\em unstable} against small$ \Delta_{V} $ and $ \Delta_{A} $.

\underline{ $ \Delta_{V} \neq 0 $ }

  There exists a line of fixed points given by  
  $ \Delta_{V}= \pi\frac{(1+w)^{3}}{1-w} \frac{\beta^{p}(w)}{w} $
  which approaches $ \infty $ as $ w \rightarrow 1^{-} $. Like the one
  loop result, $ \nu=1 $ and the flow trajectory is given by 
  $ \Delta_{V}=Cw^{2} $ (Fig.17b).  Again, we suspect that
  $ \nu=1 $ is {\em exact}
  
  The lower part of this line (thin part)
  is {\em unstable}, the higher part of this line (thick part) is
  {\em stable}. The system either flows
  to the origin or flows to the higher part depending on the initial
  condition.
  
   This line is {\em unstable} against  small $ \Delta_{M} $ and $ \Delta_{A} $.

\underline{ $ \Delta_{A} \neq 0 $ }

  The system flows to a line of fixed points given by  
  $ \Delta_{A}= \frac{\pi}{2} \frac{\beta^{p}(w)}{w} $.
  Like one loop result, the flow trajectory is given by 
  $ \Delta_{A}=C $ (Fig.17c). However,
  {\em unlike} one loop result, $ \nu^{-1}=1-\frac{4 \Delta_{A}}{\pi} 
  (1-\frac{2}{(1+w)^{2}}) $, if $ w > \sqrt{2}-1 \sim 0.41 $, $ \nu >1 $. 
  The left part of this line (thick part)
  is {\em stable}, the right part of this line (thin part) is
  {\em unstable}. At weak disorder, the system either flows
  to the left part of the line or to strong coupling regime
  depending on the initial condition. At strong disorder, the system
  always flows to strong coupling regime. 

   This line is {\em stable} against small $ \Delta_{M} $ and $ \Delta_{V} $
   in the range $ 1< w < 1.31 $. This stable region may control the integer
   quantum Hall transitions observed in real experimental systems \cite{bodo}.

    As shown first by Ludwig {\sl et al} \cite{lud},
    the random gauge fixed line is
    unstable against $ \Delta_{M} $ and $ \Delta_{V} $ 
    ( see also Eq.\ref{beta1} ). Due to the Coulomb interaction, we find
    there is a  small part of the fixed line $ 1 < w < 1.31 $ which
    is {\em stable} against small $ \Delta_{M} $ and $ \Delta_{V} $.

  From Eq.\ref{zbeta}, it
  is easy to see that $ z=1 $ on all these line of fixed points.

   Unfortunately, we are still unable to find a generic fixed points
   with all the couplings {\em non-vanishing}. These generic fixed
   points may be either unaccessible to the method developed in this paper,
   or  simply do not exist in the real experimental system.

\subsubsection{Fractional Quantum Hall Transition ($ \phi >0 $) }

Turning to the fractional case with a non-zero $\theta$, we start with
the simplest case $ w=\Delta_{V}=\Delta_{A}=0 $, $ \beta(\Delta_{M}) $
simplifies to
\begin{equation}
  \beta(\Delta_{M}) = -2 \Delta_{M}(\nu^{-1}_{p}-1)
     +\frac{2 }{\pi} \Delta^{2}_{M} 
\end{equation}
 Where $ \nu_{p} $ is the exponent in the absence of Coulomb
 interactions ($w=0$):
\begin{equation}
\nu_{p}=1-\frac{ 512 \phi (1-2 \phi) }{ N 3 \pi^{2} \lambda^{3} }
\end{equation}

   When $ 0< \phi < 1/2 $, $ \nu_{p} < 1 $, the pure fixed point is unstable,
   the system flows to a line of
   fixed points given by  
  $ \Delta_{M}= \frac{ 512 \phi (1-2 \phi) }{ N 3 \pi \lambda^{3} } $.
  From Eq.\ref{zbeta} and the fact $ \frac{\beta^{p}(w)}{w}|_{w=0}=0 $,
  we obtain $ z=1+ \Delta_{M}/\pi >1 $ which continuously changes
  along this line (Fig.18).
  From Eq.\ref{dis2}, we get $ \nu=1 $ along this line.  When $ \phi >1/2 $,
  the pure line of fixed points with $ z=1 $ is stable.

   It is easy to see that this fixed line is unstable against small
   $ (w, \Delta_{V}, \Delta_{A} ) $.

   Unlike the Coulomb interaction case ( $ \theta=0 $ ), if $ \Delta_{V}
   \neq 0 $ or $ \Delta_{A} \neq 0 $, then all the three disorders 
   are generated, this can be easily realized from Eq.\ref{beta3}.
  So we have to investigate the generic fixed points of Eq.\ref{beta3}.

   From Eq.\ref{beta3}, 
   $\beta(\Delta_{A})=0 $ implies either $ P(\tilde{\Delta}_{V})=0 $
  or $ \Delta_{M}=0 $. In the following, we discuss the two cases
  separately.

\underline{ $ P(\tilde{\Delta}_{V})=0 $ }

  From Eq.\ref{vw1} and Eq.\ref{beta3},
  $ \beta(w)=\beta(\Delta_{V})=0 $ implies that $ P(\tilde{\Delta}_{A})=0,
   \beta^{p}(w)/w= \Delta_{M}/\pi $. Finally, $ \beta(\Delta_{M})=0 $ implies 
\begin{equation}
   -2 \Delta_{M}(\tilde{\nu}^{-1}_{p}-1)
     +\frac{1 }{\pi} \tilde{\Delta}^{2}_{C}  
     +\frac{4 }{\pi} \tilde{\Delta}_{V} \tilde{\Delta}_{A} 
     -\frac{2 }{\pi} \tilde{\Delta}_{C} P( \tilde{\Delta}_{C} ) 
     -\frac{16}{\pi} \Delta_{M} P(\tilde{\Delta}_{K})=0 
\label{bad}
\end{equation}

  The {\em disordered} critical exponent Eq.\ref{point} is simplified to
\begin{equation}
   \nu^{-1}= \tilde{\nu}^{-1}_{p} + \frac{ 8 P(\tilde{\Delta}_{K})}{\pi} 
\label{dis3}
\end{equation}

     It is easy to see that $ P(\tilde{\Delta}_{V})=
     P(\tilde{\Delta}_{A})=0 $ implies the following equation:
\begin{equation}
     t = \frac{\Delta_{V}}{\Delta_{A}}=
     \frac{ \phi (3+w-\phi) }{ w+3 \phi -1} = 
     \frac{(1+w)[(3+w) \phi-(1+w)^{2}]}{\phi (3+w-\phi)}
\end{equation}

     Namely, $ x=1+w $ should satisfy the {\em fourth} order equation:
\begin{equation}
   x^{4} +2 (\phi-1) x^{3} -2 \phi^{2} x^{2} -2 \phi (\phi-1)(\phi+2) x
   + \phi^{2} (2-\phi)^{2} =0
\label{root}
\end{equation}
    with the constraints $ 3+w > \phi, \phi > \frac{1-w}{3},
    \phi > \frac{(1+w)^{2}}{3+w} $.

   From Ref.\cite{jinwu}, $ \beta^{p}(w)/w= \Delta_{M}/\pi > 0 $ 
   implies that $ \phi < 2 $, therefore $ w < \sqrt{5} $. 

  If $ \phi=0 $, Eq.\ref{root} reduces to $ x^{3}(x-2)=0 $, therefore $ w=1 $.
  If $ 0< \phi < 1 $, there is {\em no} real root which satisfies both
  $ x > 1 $ and the constraints.

  If $ \phi=1 $, Eq.\ref{root} reduces to $ (x^{2}-1)^{2}=0 $ which implies
  that $ w=0 $. Substituting $ (\phi=1, w=0, t=1) $ into Eq.\ref{pfun},
  we find $ \tilde{\Delta}_{C}= P(\tilde{\Delta}_{C})=
  P(\tilde{\Delta}_{K})=0 $.

   If $ 1< \phi < 2 $, there is only one real root with $ x > 1 $.
   We also find $ P(\tilde{\Delta}_{K}) < 0 $,
  therefore $ \nu > \tilde{\nu}_{p} > 1 $ in this regime.

  Unfortunately, when substituting the $ ( \phi, w, t ) $ into Eq.\ref{bad},
  we find that the left hand side of the equation is always {\em positive}.
  Therefore, we conclude there is {\em no} perturbatively accessible
  fixed points with $ \Delta_{M}, \Delta_{V}, \Delta_{A} > 0 $.

\underline{ $ \Delta_{M}=0 $ }

    From Eq.\ref{vw1}, we see that $\beta(w)=0 $ implies 
    $ \beta(\Delta_{V})=0 $. From Eq.\ref{beta3}, it leads to
\begin{equation}
  \frac{ \beta^{p}(w)}{w}  = \frac{1}{\pi}( P(\tilde{\Delta}_{V})
	       + 2 P( \tilde{\Delta}_{A}) )
\label{fix1}
\end{equation}    

    From  Eq.\ref{beta3},  $\beta(\Delta_{M})=0 $ leads to
\begin{equation}
     \frac{1 }{\pi} \tilde{\Delta}^{2}_{C}  
     +\frac{4 }{\pi} \tilde{\Delta}_{V} \tilde{\Delta}_{A}  
     -  \frac{4 }{\pi} \tilde{\Delta}_{A}  P( \tilde{\Delta}_{V} )  
     -  \frac{4 }{\pi} \tilde{\Delta}_{V} P( \tilde{\Delta}_{A} ) 
     -\frac{2 }{\pi} \tilde{\Delta}_{C} P( \tilde{\Delta}_{C} ) =0
\label{fix2}
\end{equation}
      
   From the last equation, given $ ( \phi, t ) $, we can determine $ w $,
   then substituting $ (\phi, w, t ) $ to Eq.\ref{fix1}, we can determine
   $ \Delta_{A}, \Delta_{V} $. In the following, we consider the two end
    lines separately:

 (a) $ \Delta_{V}=0 $ ( namely $ t=0 $ )

 If $ \phi=0 $, Eq.\ref{fix2} becomes an identity,
 we recover the results of the IQHT (Fig.17c).

 If $ \phi > 0 $, the solution of Eq.\ref{fix2} is $ w=\frac{1}{2}
 [ 3 \phi +\sqrt{ (3 \phi)^{2}+4(1-\phi) }-4] > 0 $ (namely, 
 $ \phi > 3/5 $ ). Substituting this
 expression into Eq.\ref{fix1}, we can determine $ \Delta_{A} $ and
 the constraint  $ \phi < 1.3 $. Thus $ \phi $ must satisfy $ 3/5 < \phi < 1.3 $.
  From Eq.\ref{point}, we find this line is stable against small $ \Delta_{M} $.

 (b) $ \Delta_{A}=0 $ ( namely $ t=\infty $ )

 If $ \phi=0 $, Eq.\ref{fix2} becomes an identity,
 we recover the results of the IQHT (Fig.17b).

 If $ \phi > 0 $, the solution of Eq.\ref{fix2} is $ w= 3-5 \phi $.
 Substituting $ w=3-5 \phi $ into Eq.\ref{fix1}, we can
 determine $ \Delta_{V} $:
 $ \frac{\beta^{p}(w) }{w}= 
 \frac{\Delta_{V}}{ 32 \pi} \frac{6 \phi-1}{ (1-\phi)^{2} } $.
 Therefore $ \phi $ must satisfy $ 1/6 < \phi < 3/5 $.
 Eq.\ref{point} becomes $ \nu^{-1}= \tilde{\nu}^{-1}_{p} + \frac{ 3 \Delta_{V}}
 {8 \pi} \frac{ \phi^{2}}{ (1-\phi)^{3} } > \tilde{\nu}^{-1}_{p} > 1 $,
  therefore, this line is unstable against small $ \Delta_{M} $.
 
    We conjecture that there is a fixed {\em plane } 
    which connect the above two end lines at
   $ t=0 $ and at $ t=\infty $ (Fig.19). The shaded (unshaded) region
   has $ \nu > 1 ( \nu <1 ) $, therefore
   is stable (unstable) against small $ \Delta_{M} $. Numerical
   analysis is needed to determine its precise boundary. The stable
   region may control the fractional quantum hall transitions observed
   in real experimental systems \cite{shahar}.

\section{Conclusion}

   Recent experiments indicated that the transitions between two
   Quantum Hall states or between a quantum Hall state and an
   insulating state may be described by quantum critical theories.
  In these theories, different FQH states and insulating states
  are considered as different ground states of the electron systems.
  The three important questions remain unsolved on the nature of these quantum
   phase transitions are:

   (a) The effects of the quasi-particle statistics 

   (b) The effects of long-ranged Coulomb interaction on the transitions 

   (c) The effects of all kinds of disorders

    Answering the three questions at the same time seems
    a forbidding task at this moment. Ref.\cite{jinwu} investigated
    the combined effects of (a) and (b) in a Dirac fermion model and found a line
    of fixed points.  Along this line, both Chern-Simon interaction and Coulomb
    interaction are non-vanishing, the dynamic exponent $ z=1 $.
    In this paper, we make a serious
    attempt to study the combined effects of (a), (b) and (c) in the
    Dirac fermion model.
    We perform a renormalization Group analysis
    by the systematic perturbative expansions in $ 1/N $ ( $ N $ is the No.
    of species of Dirac fermions) and the variances of three disorders
    $ \Delta_{M}, \Delta_{V}, \Delta_{A} $.
  we find that $ \Delta_{M} $ is irrelevant along this line;
  there is {\em no} $1/N $ correction to $ \beta(\Delta_{A}) $,
  therefore $ \Delta_{A} $ is always marginal;
   $ \Delta_{V} $ is marginal along this line,
  , irrelevant {\em above} this line, relevant {\em below} this line.
  With the presence of all the three disorders, the pure fixed line is 
  unstable.

    In IQHT, in the three special cases,
    we find the three non-trivial lines of fixed points
    with dynamic exponent $ z=1 $.
    
    The fixed line in $ (\Delta_{M}, w ) $
    plane has $ \nu=1 $ and is unstable against small
    $ (\Delta_{V}, \Delta_{M} ) $.

     The fixed line in $ (\Delta_{V}, w ) $
    plane has $ \nu=1 $ and is unstable against small
    $ (\Delta_{M}, \Delta_{A} ) $.

      Most interestingly, the fixed line in $ (\Delta_{A}, w ) $
    plane has continuously changing $ \nu $ 
    and is {\em stable} against small $ (\Delta_{M}, \Delta_{V} ) $ in
    the small range $ 1<w < 1.31 $ (Fig.17),
   this stable region may control the integer
   quantum Hall transitions observed in real experimental systems \cite{bodo}.

    The results may be relevant to
    the IQH to insulator transitions. It may also be important to
     high $ T_{c} $ superconductors. Because it was well-established that
    high $ T_{c} $ superconductors have a d-wave order parameter
    and its quasiparticle excitations are described by $ 2+1 $ dimensional
    Dirac fermions \cite{senthil}.

        In FQHT, setting Coulomb interaction to zero and
	$ \Delta_{V}=\Delta_{A}=0 $,
	we find a line of fixed points with $ \nu=1 $ and
	$ z >1 $ which continuously
	changes along this line (see Fig.18). This line is unstable
	against small $ (w, \Delta_{V}, \Delta_{A}) $.

        Most interestingly, setting $ \Delta_{M} =0 $,
	we find a fixed plane with $ z=1 $,
	 the part of this plane with $ \nu > 1 $ 
	is stable against small $ \Delta_{M} $.
   This stable region may control the fractional
   quantum Hall transitions observed in real experimental systems \cite{shahar}.

   Unfortunately, we are unable to find a generic fixed points
   with all the couplings {\em non-vanishing}. These generic fixed
   points may be either unaccessible to the method developed in this paper,
   or  simply do not exist in the real experimental system.
   However, by looking at carefully the divergent structures of all the relevant
      Feymann diagrams, we show the model is {\em renormalizable}
      to the order $ (1/N) \Delta, \Delta^{2}, (1/N)^{2} $ discussed
      in this paper; we do bring out the
      systematic and elegant structure which describes the interferences
      between Chern-Simon interaction, Coulomb interaction and the three
      kinds of disorders. We believe that the structure is interesting 
      in its own right and may inspire
      future work to study this difficult problem.

\centerline{\bf Acknowledgments}
We thank M. P. A. Fisher, B.~Halperin, A. Millis, C. Mudry, S.~Kivelson,
N.~Read, S. Sachdev and X. G. Wen for helpful discussions.
This work was supported by NSF Grant No. DMR-97-07701.

\appendix

\section{ The proof of equivalence of random vector gauge potential
    $ \Delta_{A} \delta_{ij} $ and $\Delta_{A} (\delta_{ij}-\frac{q_{i} q_{j}}
    {q^{2}} ) $.}

  We can decompose the random gauge field $ A_{i}(x) $ in Eq.\ref{classical}
  into transverse and longitudinal components:
\begin{equation}
  A_{i}= A^{T}_{i}+A^{L}_{i}=
  \epsilon_{ij} \partial_{j} \chi^{T}+\partial_{i} \chi^{L}
\end{equation}
   Where $ \chi^{T}(x), \chi^{L}(x) $ satisfy
\begin{equation}
   <\chi^{T}(x) \chi^{T}(x^{\prime})>=
   <\chi^{L}(x) \chi^{L}(x^{\prime})>= -\Delta_{A} \ln |x-x^{\prime}|
\end{equation}

    From the above equation, it can be shown easily
\begin{eqnarray}
  < A^{T}_{i}(k) A^{T}_{j}(k^{\prime}) > & = &\Delta_{A} 
    ( \delta_{ij} - \frac{k_{i} k_{j}}{ k^{2}} ), \nonumber   \\
  < A^{L}_{i}(k) A^{L}_{j}(k^{\prime})> &  = &  \Delta_{A} 
     \frac{k_{i} k_{j}}{ k^{2}}
\end{eqnarray}

     Adding the two equations above leads to 
\begin{equation}
   < A_{i}(x) A_{j} (x^{\prime} ) >=
   < A^{T}_{i}(x) A^{T}_{j} (x^{\prime} ) >
   + < A^{L}_{i}(x) A^{L}_{j} (x^{\prime} ) >
   = \Delta_{A} \delta_{ij} \delta^{d}(x-x^{\prime})
\end{equation}
  which is the third equation of Eq.\ref{aver}.
   
   By gauge transformation, $ A^{L}_{i}(x)= \partial_{i} \chi^{L}(x) $
   can be removed, so $ 
  < A^{L}_{i}(k) A^{L}_{j}(k^{\prime})>   =   \Delta_{A} 
     \frac{k_{i} k_{j}}{ k^{2}} $ should {\em not } make any contribution
     to {\em  gauge-invariant} quantities. This fact is similar to
     the "running gauge fixing parameter " in usual relativistic
     quantum field theory \cite{ryder}. This point can also be demonstrated
     in  the following specific example.

    Let us evaluate its contribution to fermion self energy Fig.4c, the 
    divergent part is
\begin{equation}
  \frac{1}{2 \pi \epsilon} ( \gamma_{0} \omega+ \gamma_{i} k_{i} ) 
\end{equation}

    It is evident that although this contribution infect
    the anomalous dimension of
    the field operator $ \psi $, it does not affect the dynamic
     exponet $ z $ which is a gauge invariant quantity.

\vspace{1cm}

\centerline{\bf Figure Captions}

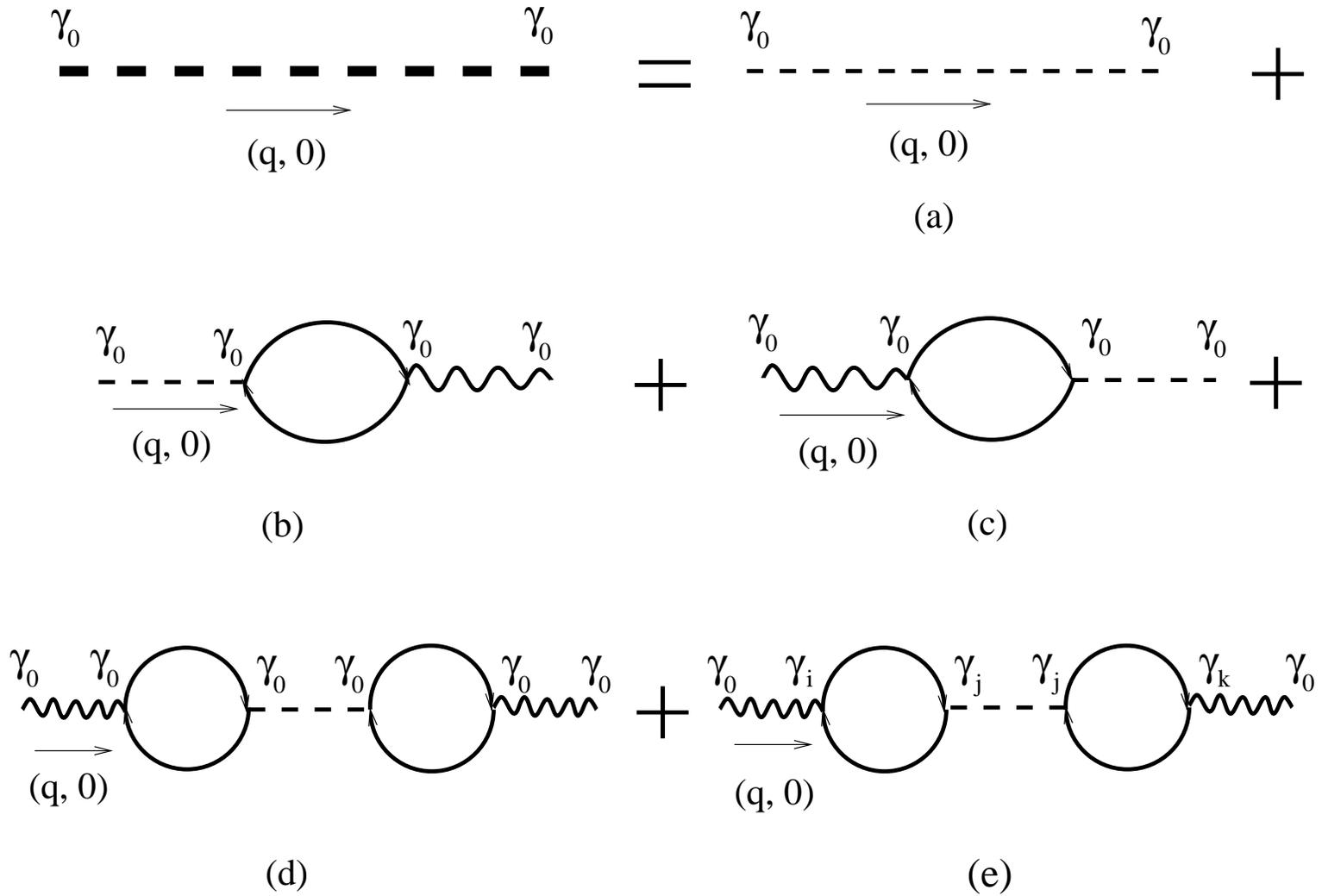
\begin{figure}
\caption{ The effective random scalar potential $ \tilde{\Delta}_{V} $
	to the order $ (1/N)^{0} \Delta_{V} $ and 
	$ (1/N)^{0} \Delta_{A} $. The thick dashed line is
	the effective random scalar potential. The thin dashed lines are
	the bare random scalar and vector potentials.
	The wave line is the gauge fields propagators.}
\label{fig.1}
\end{figure}

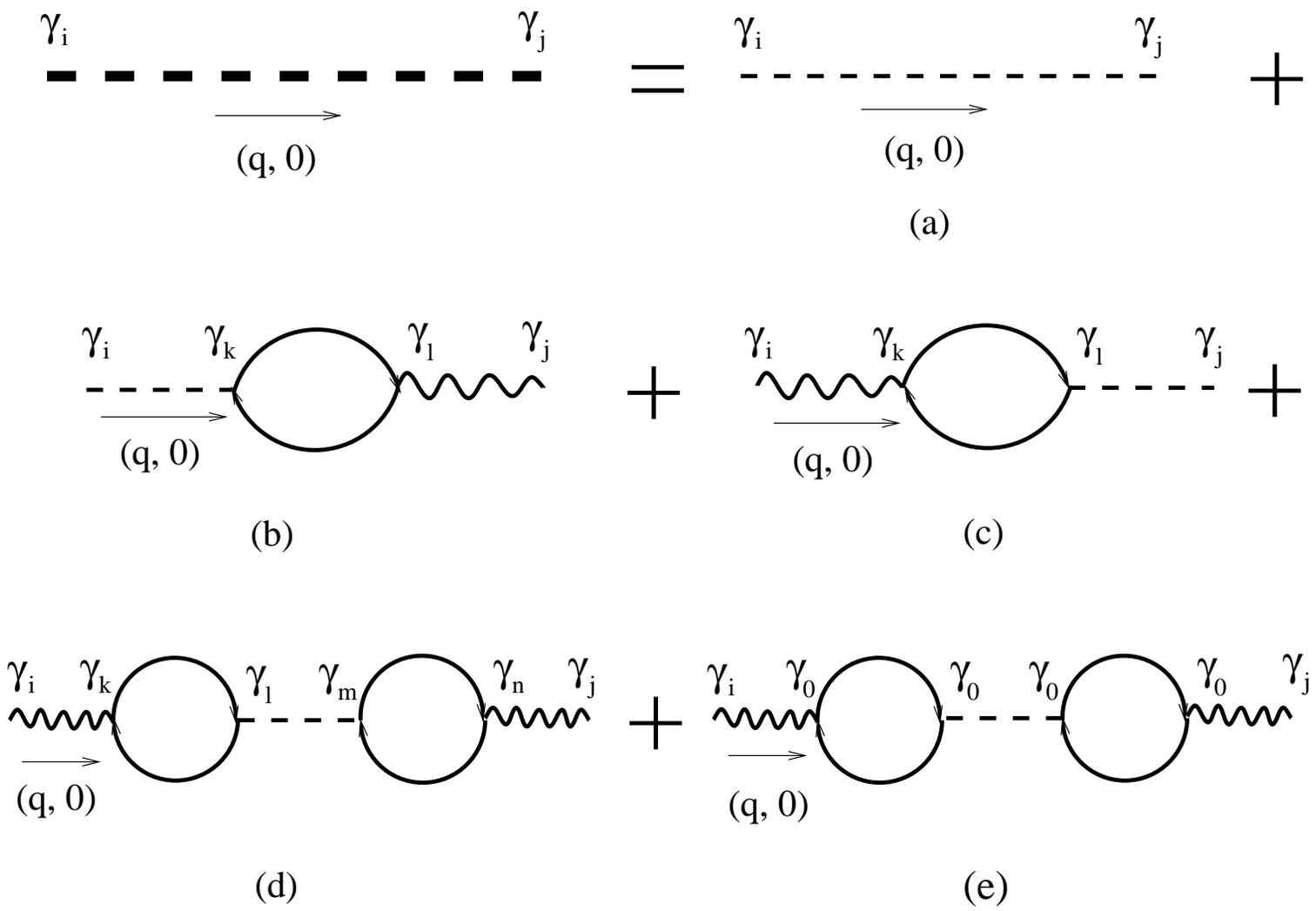
\begin{figure}
\caption{ The effective random vector potential $ \tilde{\Delta}_{A} $
	to the order $ (1/N)^{0} \Delta_{A} $ and $ (1/N)^{0} \Delta_{V} $.
	The thick dashed line is
	the effective random vector potential. The thin dashed lines are
	the bare random scalar and vector potentials.}
\label{fig.2}
\end{figure}

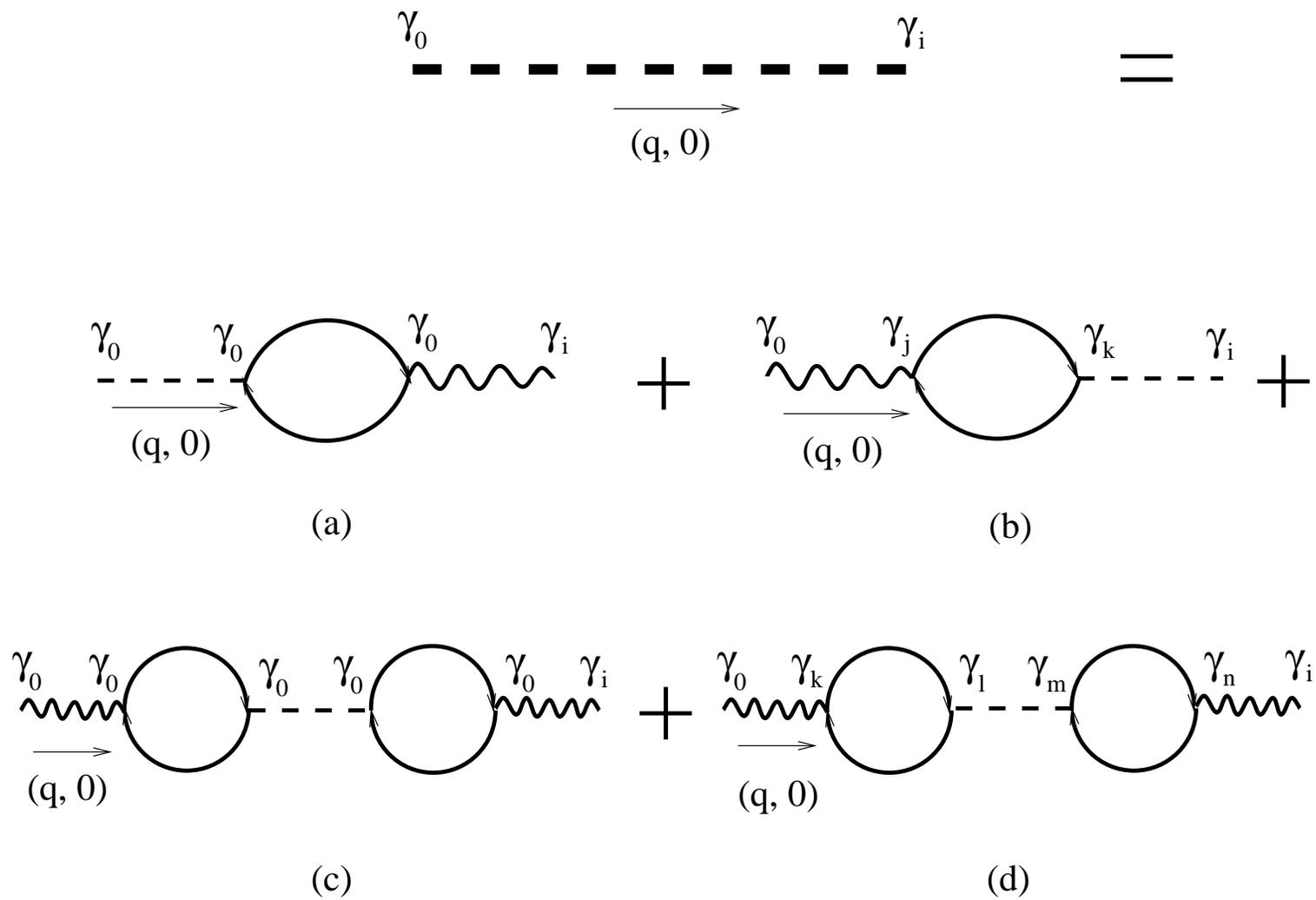
\begin{figure}
\caption{ The effective random Scalar-Vector potential $ \tilde{\Delta}_{C} $
	to the order $ (1/N)^{0} \Delta_{A} $ and $ (1/N)^{0} \Delta_{V} $.
	The thick dashed line is the effective random SV potential. There is
	{\em no} bare random SV potential.}
\label{fig.3}
\end{figure}

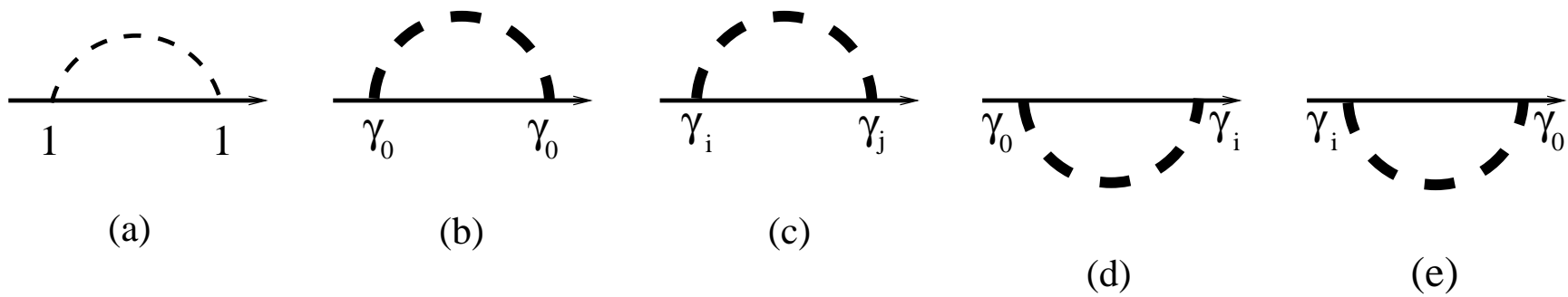
\begin{figure}
\caption{ The contribution to the Dirac fermion self-energy from
	random potentials.}
\label{fig.4}
\end{figure}

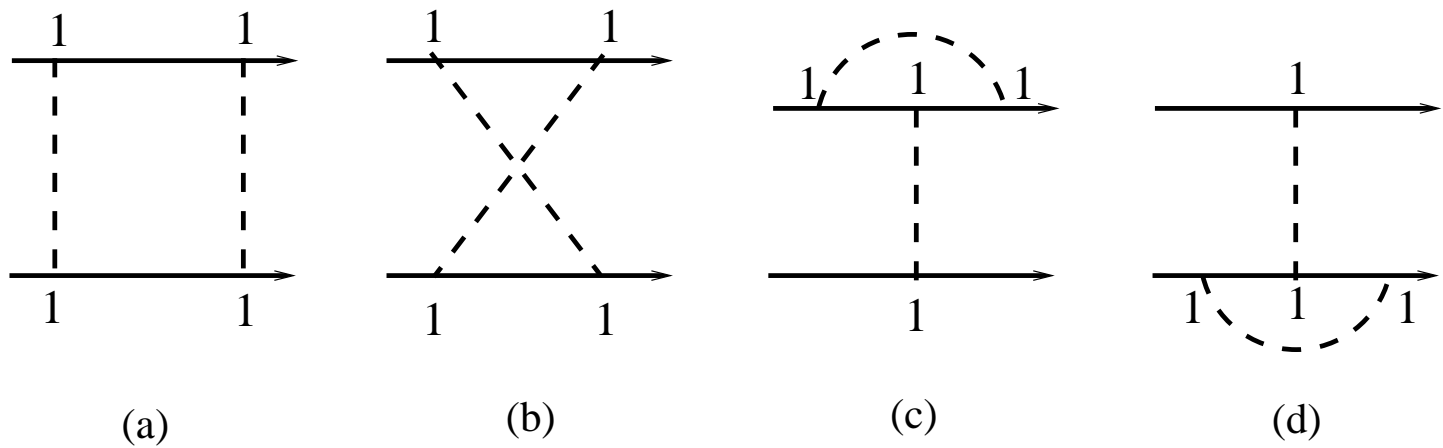
\begin{figure}
\caption{ The renormalization from the random mass. }
\label{fig.5}
\end{figure}

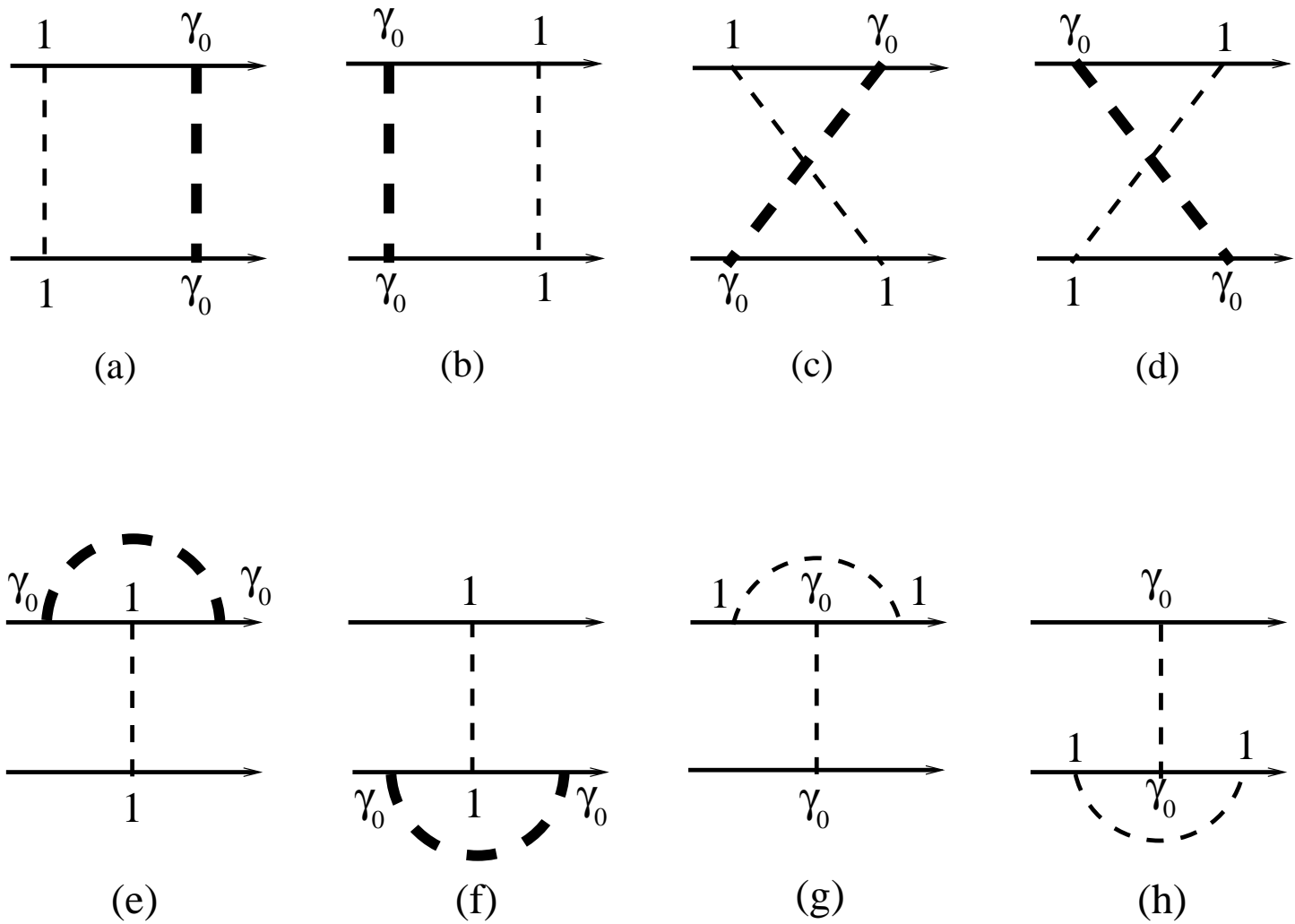
\begin{figure}
\caption{ The renormalization from the interference of random mass 
 and {\em effective} random scalar potential. Note the effective
 random scalar potential can only appear in the {\em internal loops}}
\label{fig.6}
\end{figure}

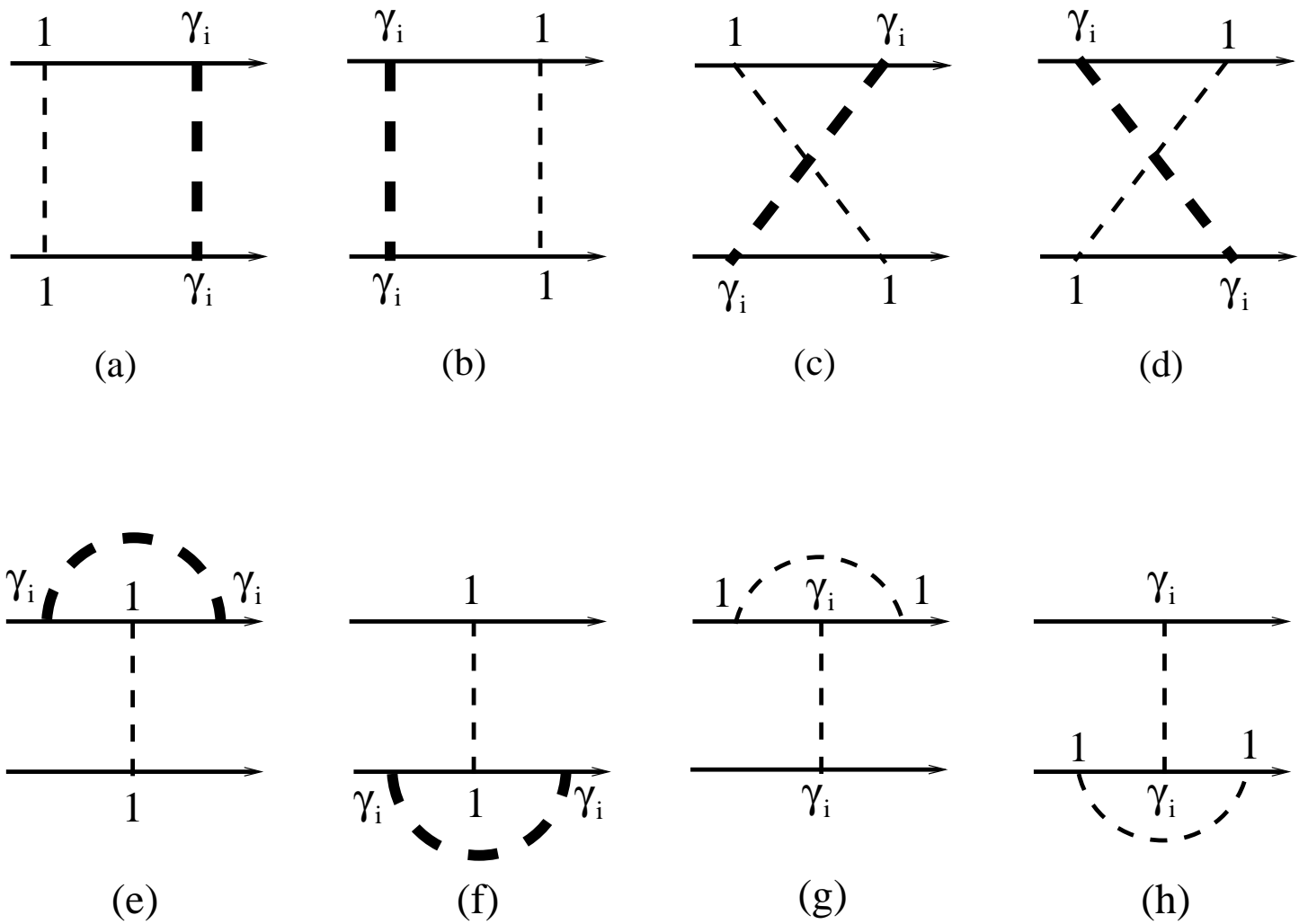
\begin{figure}
\caption{ The renormalization from the interference of random mass 
 and {\em effective} random vector potential}
\label{fig.7}
\end{figure}

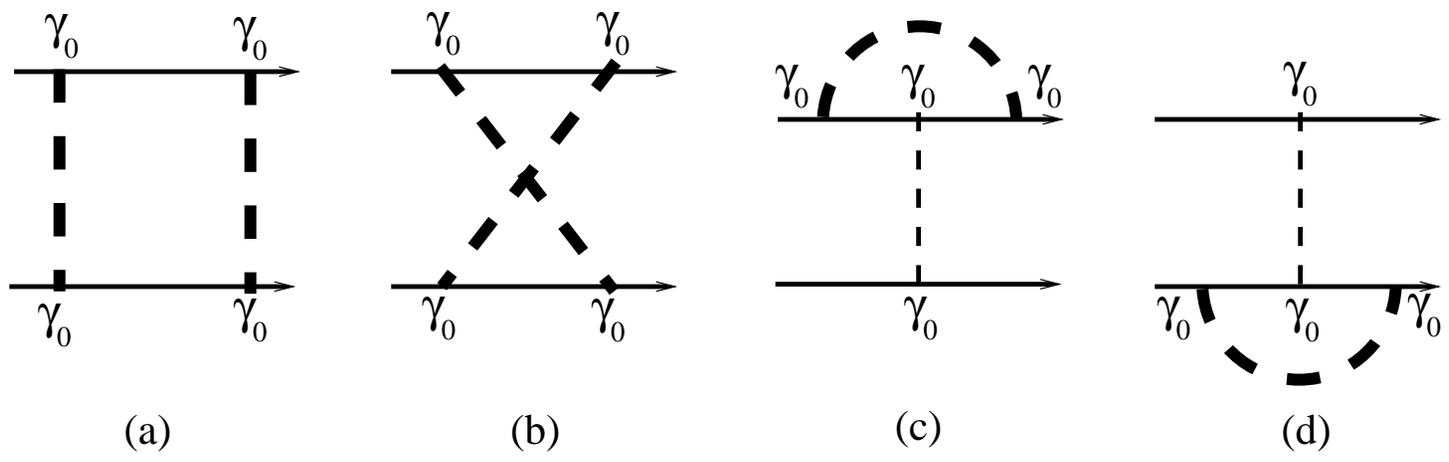
\begin{figure}
\caption{ The renormalization from the random scalar potential} 
\label{fig.8}
\end{figure}

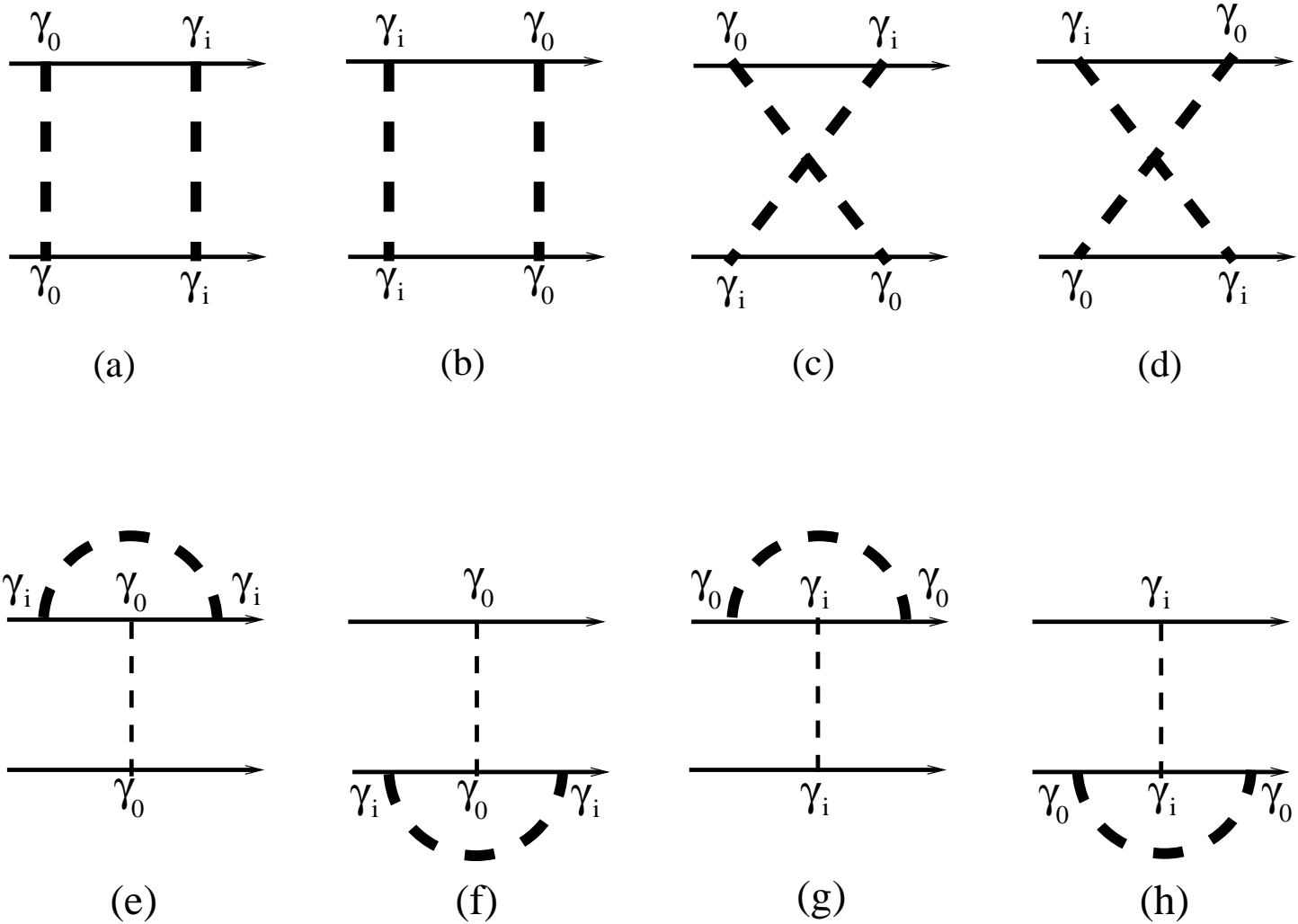
\begin{figure}
\caption{ The renormalization from the interference of effective random
 scalar potential and effective random vector potential}
\label{fig.9}
\end{figure}

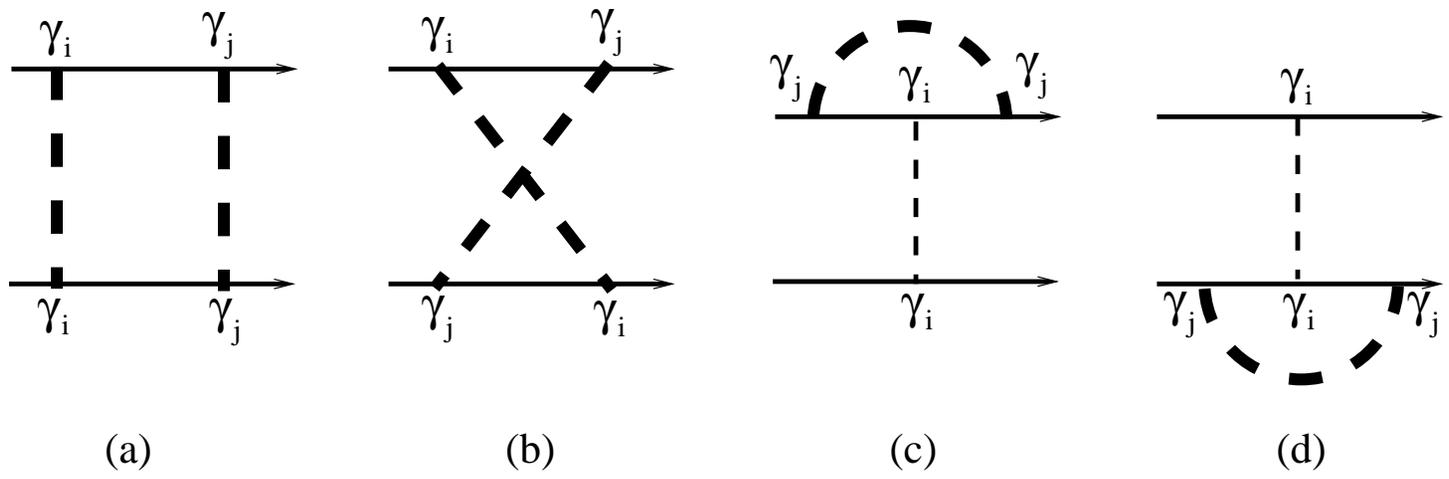
\begin{figure}
\caption{ The renormalization from the effective random vector potential} 
\label{fig.10}
\end{figure}

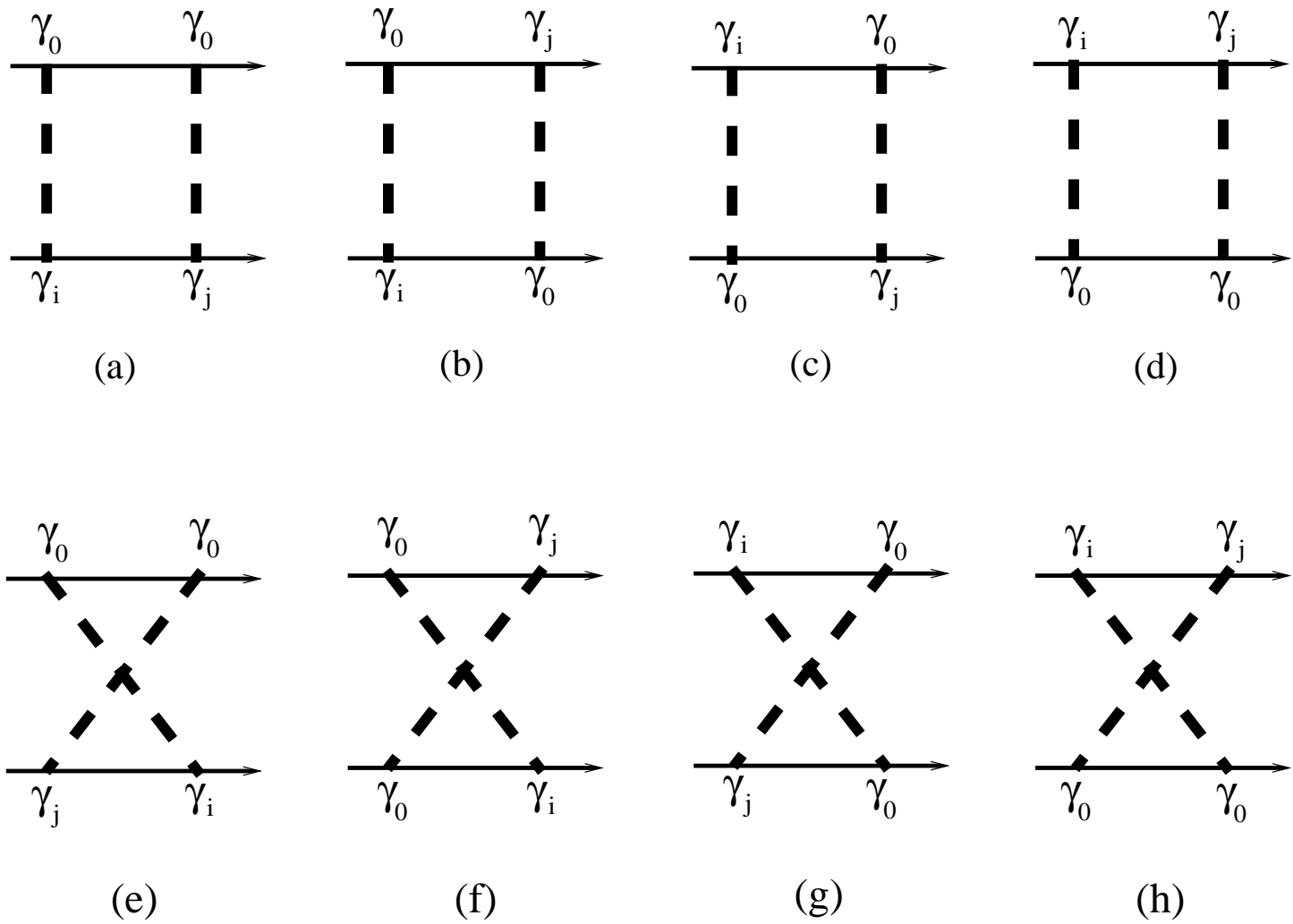
\begin{figure}
\caption{ The renormalization from the effective random SV potential} 
\label{fig.11}
\end{figure}

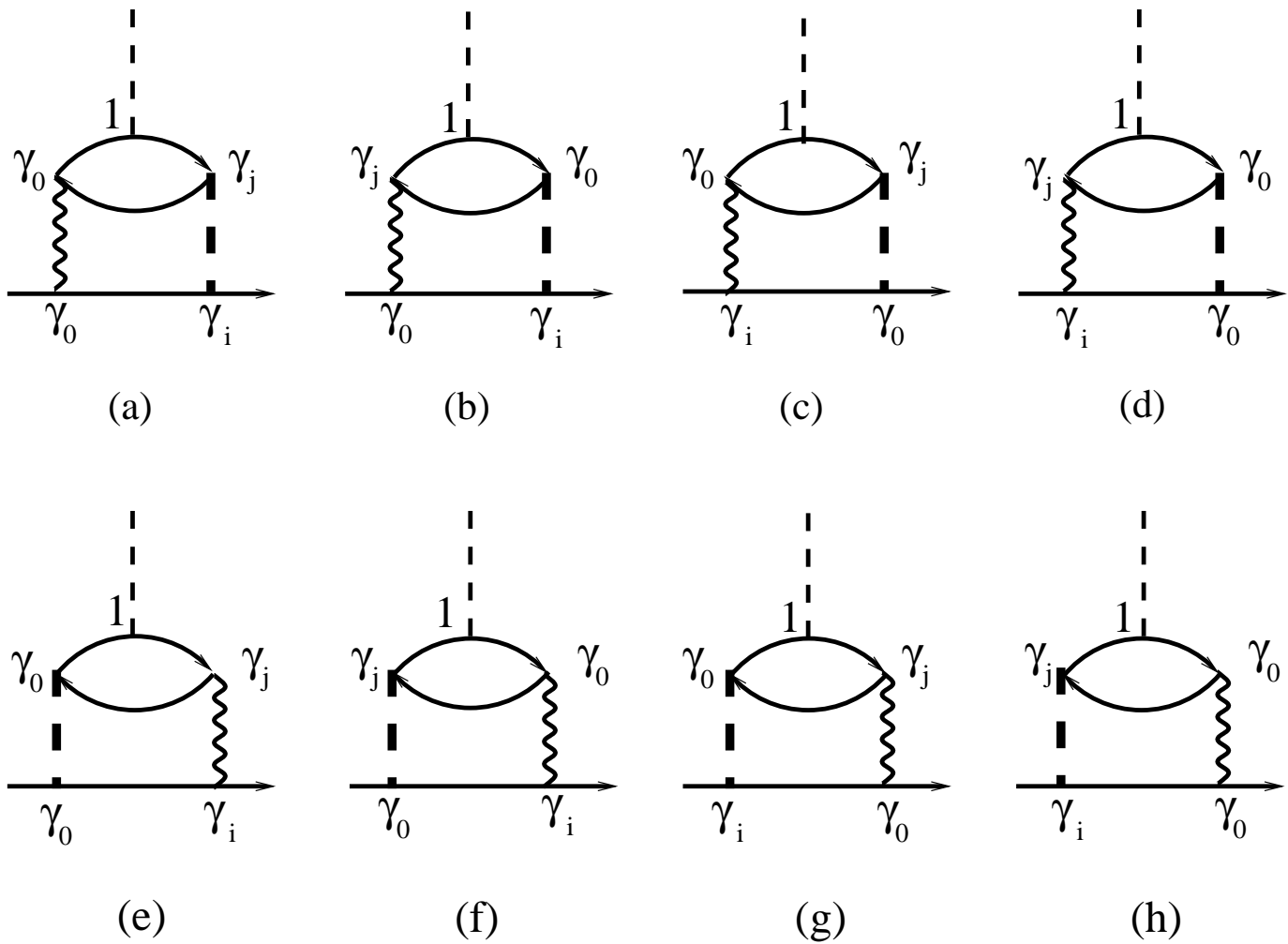
\begin{figure}
\caption{ The Renormalization to random mass from $ \tilde{\Delta}_{V},
 \tilde{\Delta}_{A}, \tilde{\Delta}_{C} $. In the text, it is called 
 $ \tilde{\Delta}_{K} $. Compare this figure to Fig.13. }
\label{fig.12}
\end{figure}

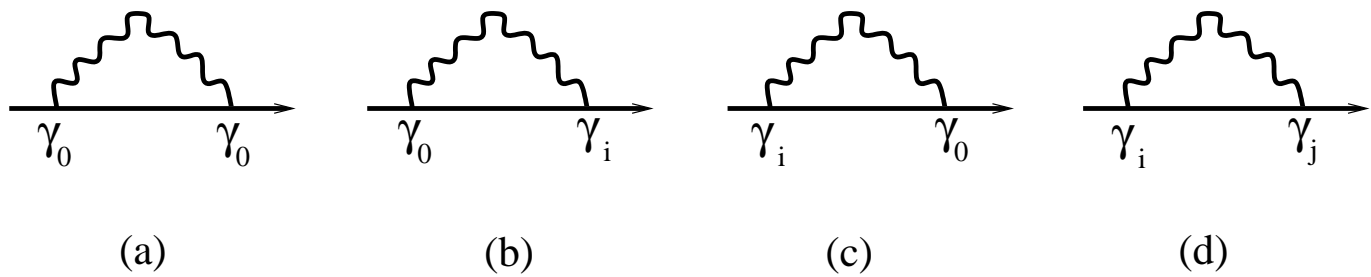
\begin{figure}
\caption{ The renormalization to the Dirac fermion self-energy from
	 gauge fields fluctuations to order $ 1/N $ }
\label{fig.13}
\end{figure}

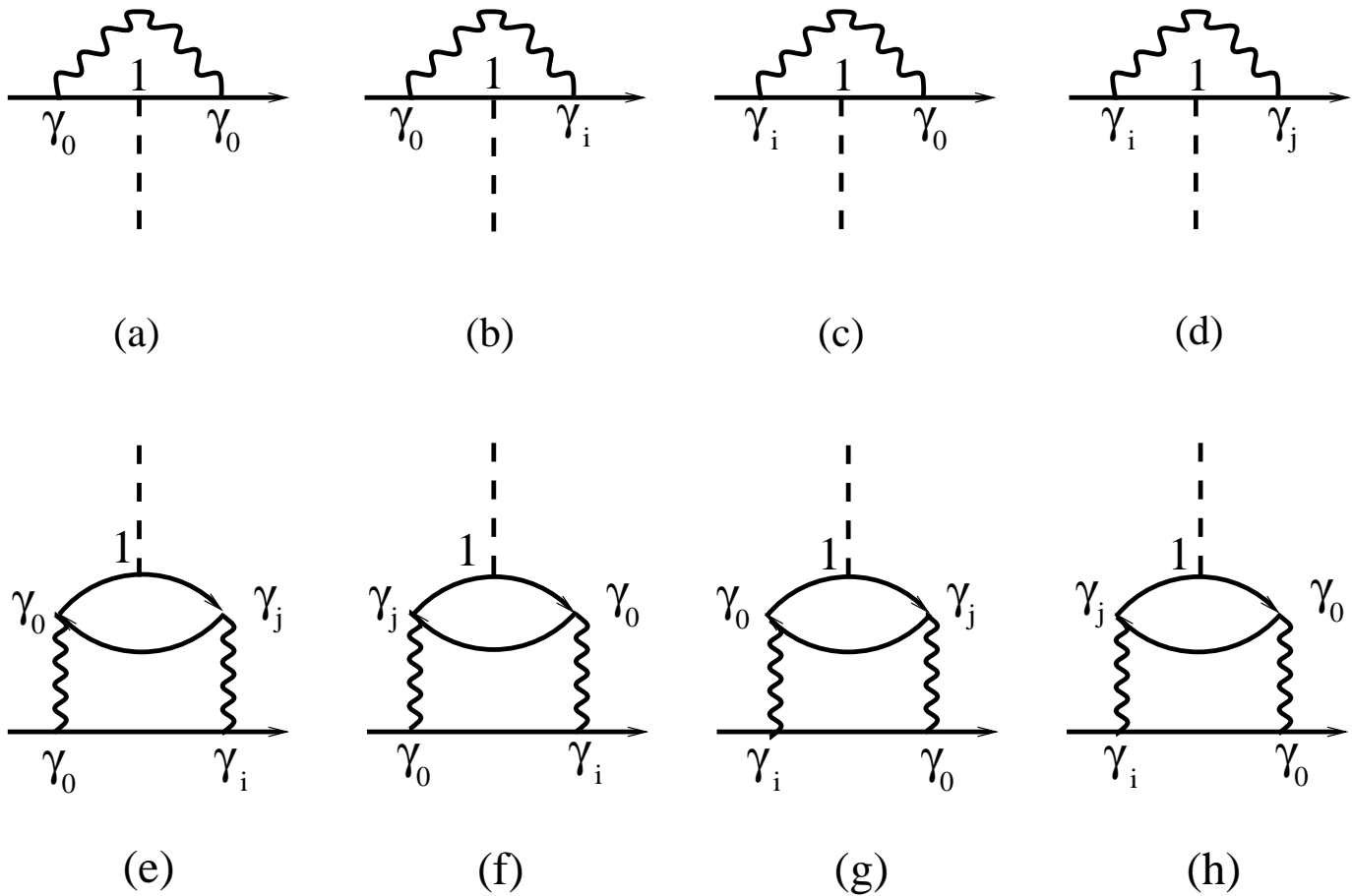
\begin{figure}
\caption{ The renormalization to the random mass from
	 gauge fields fluctuation to the order $ 1/N $ }
\label{fig.14}
\end{figure}

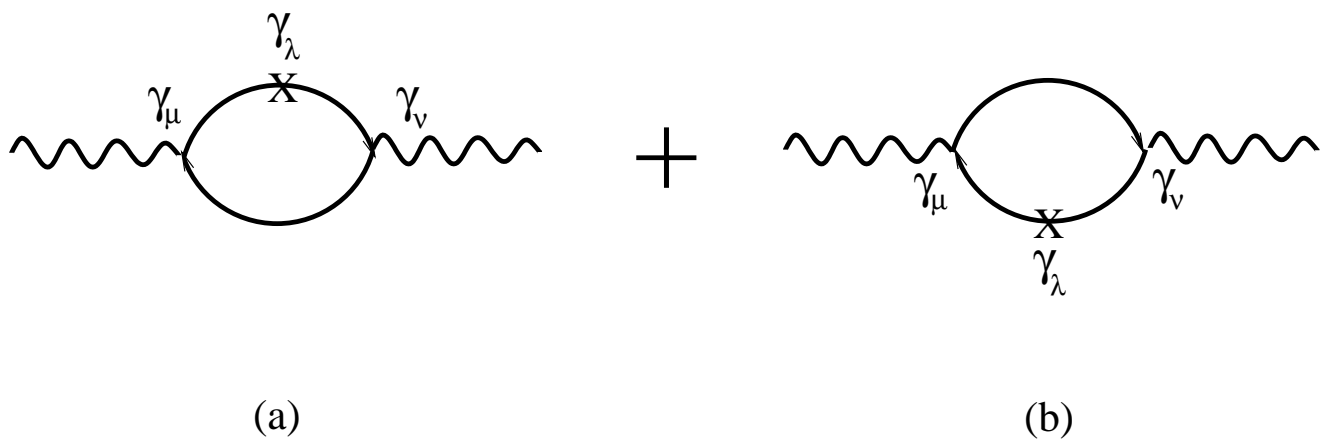
\begin{figure}
\caption{ The fermion bubbles which contribute to the
      renormalization of random scalar and vector potentials.
      In the figure, $ \mu, \nu, \lambda=0, 1 ,2 $.}
\label{fig.15}
\end{figure}

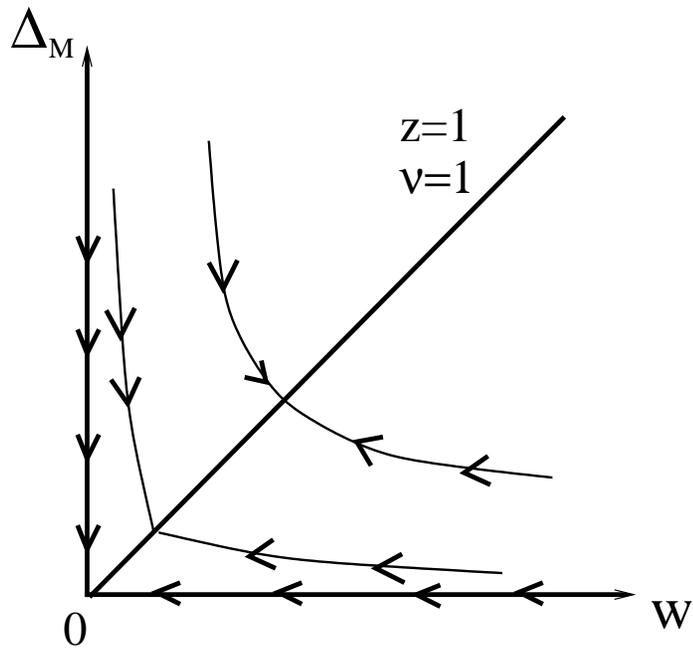
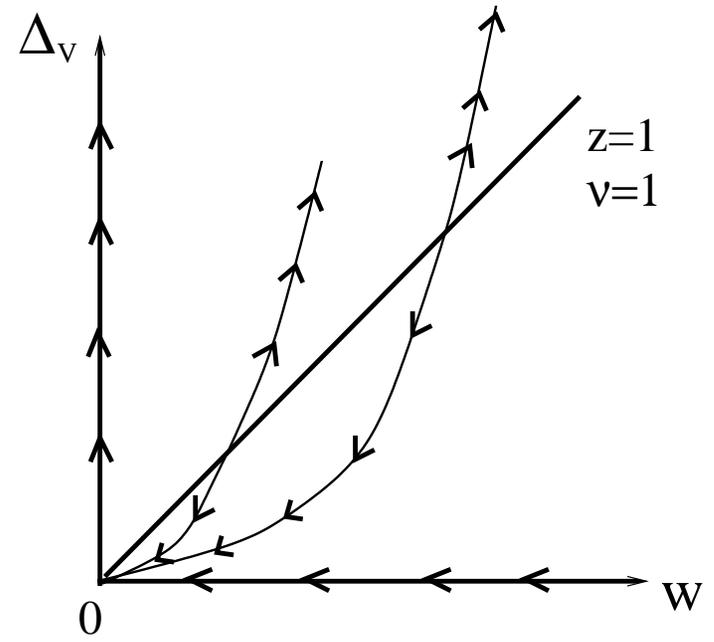
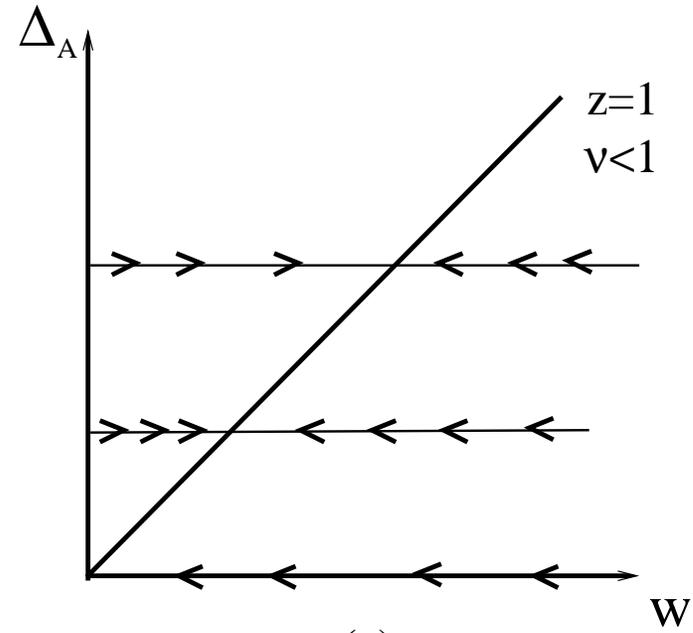
\begin{figure}
\caption{ The One-Loop Renormalization Group flow of (a)
     random mass and Coulomb interaction (b) random scalar potential
     and Coulomb interaction (c) random gauge potential and Coulomb interaction }
\label{fig.16}
\end{figure}

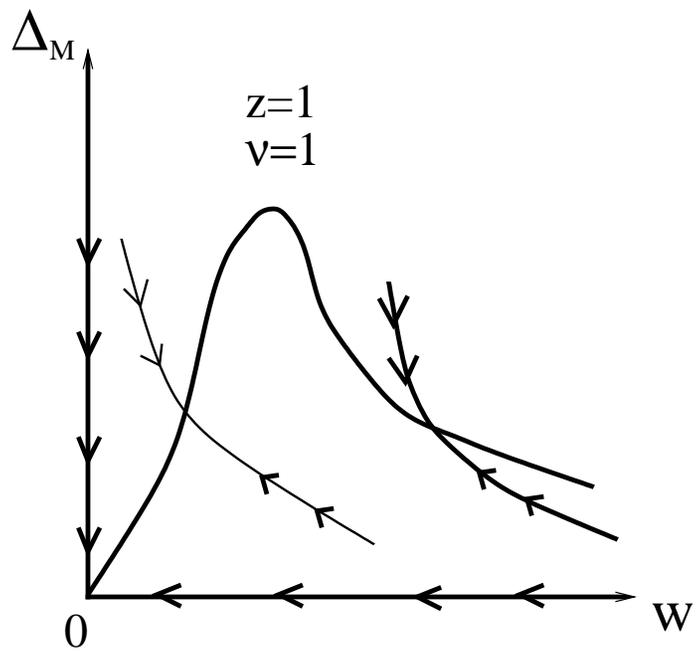
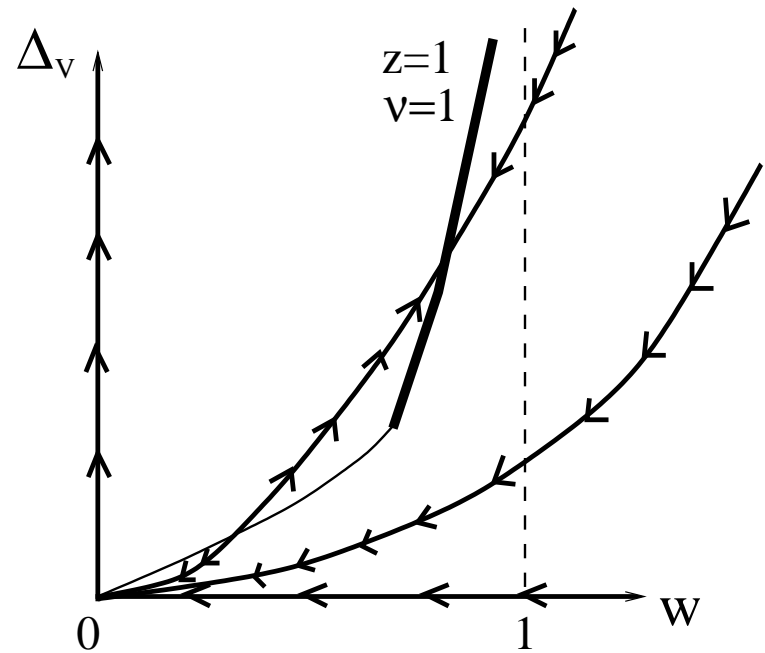
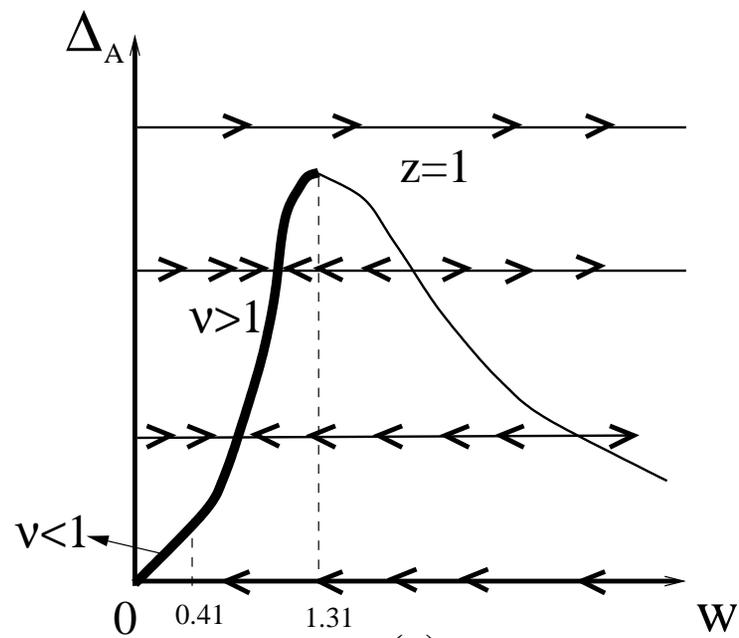
\begin{figure}
\caption{ The Renormalization Group flow to order $ 1/N $ of (a)
     random mass and Coulomb interaction (b) random scalar potential
     and Coulomb interaction (c) random gauge potential and Coulomb interaction.
     The thick (thin) lines in (b) and (c) are {\em stable}({\em unstable} )
     line of fixed points.}
\label{fig.17}
\end{figure}

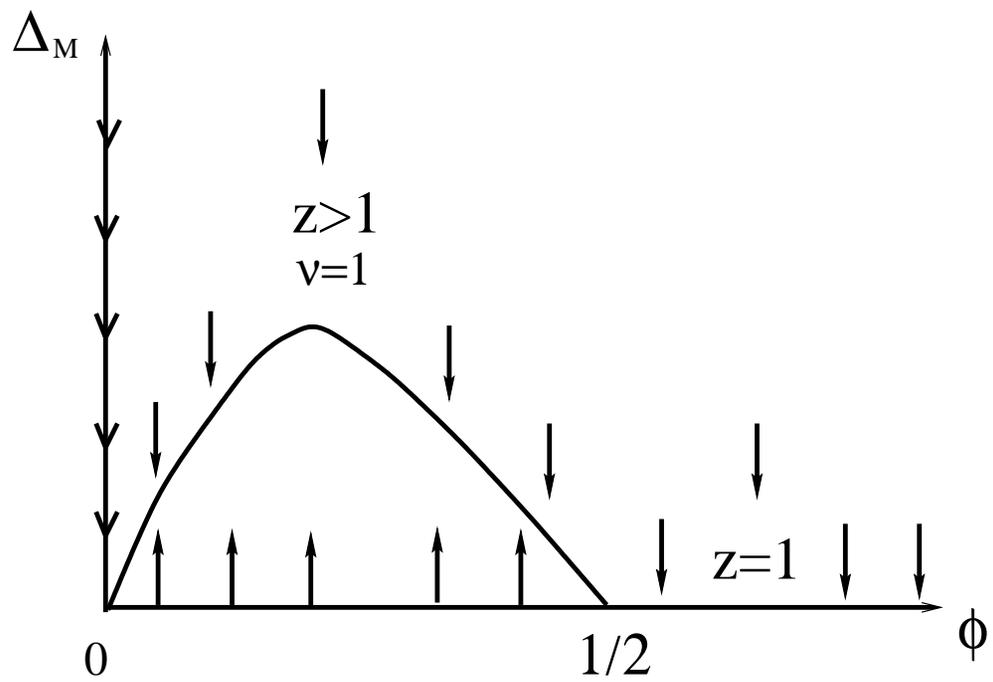
\begin{figure}
\caption{ The Renormalization Group flow to order $ 1/N $ of
     random mass and Chern-Simon interaction }
\label{fig.18}
\end{figure}

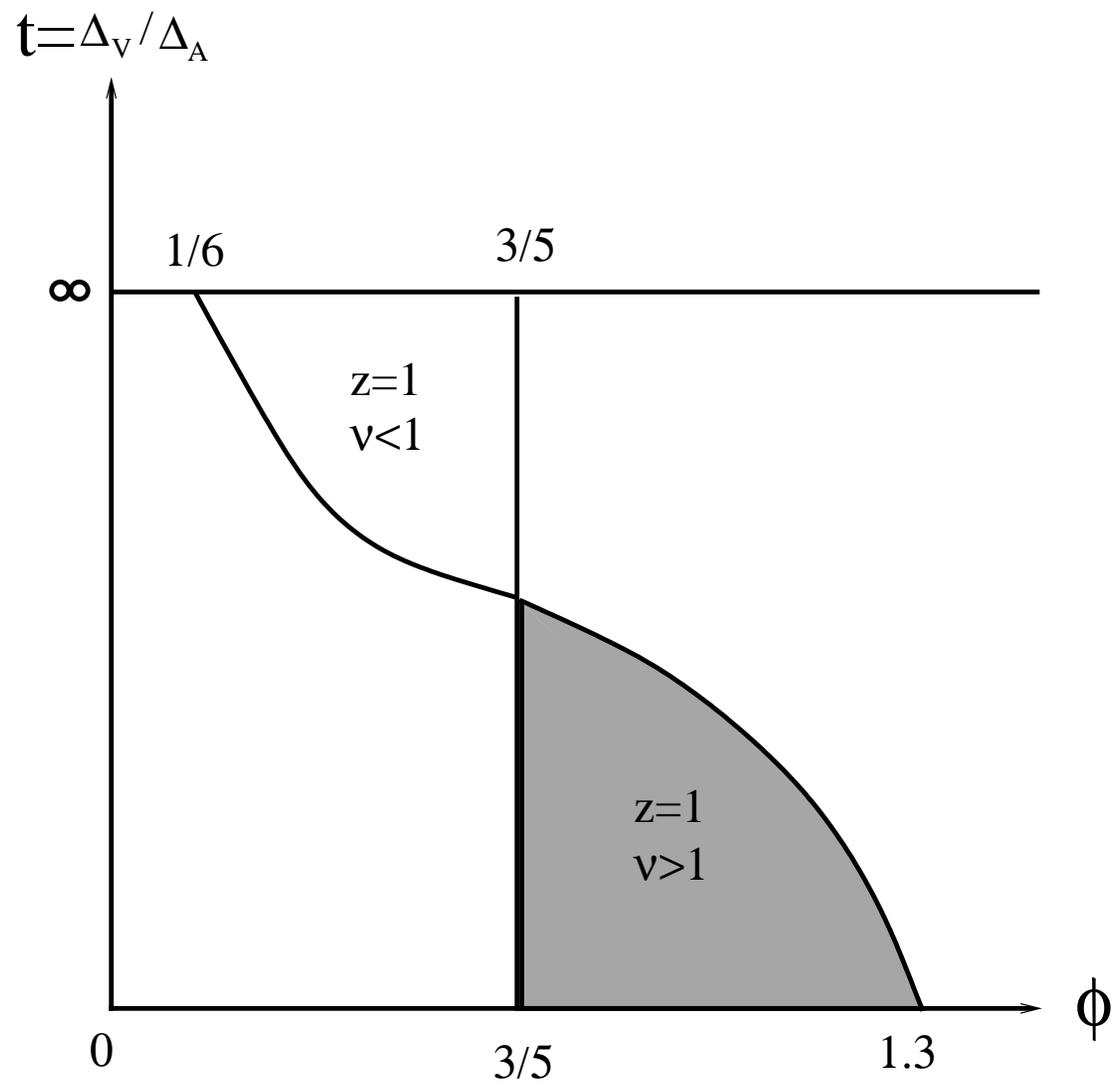
\begin{figure}
\caption{ If $ \Delta_{M}=0 $, there is a fixed plane in 
 $ ( \phi, t=\Delta_{V}/\Delta_{A} ) $ plane. $ w $ and $ \Delta_{A} $
 and $ \Delta_{V} $ are uniquely determined by $ (\phi, t ) $, therefore are not  shown in the figure.  The shaded regime with
 $ z=1, \nu > 1 $ is stable against small $ \Delta_{M} $.}
\label{fig.19}
\end{figure}

\end{document}